%% file: main.tex
\documentclass[lettersize,journal]{IEEEtran}
\usepackage{amsmath,amsfonts}
\usepackage{algorithmic}
\usepackage{algorithm}
\usepackage{array}
\usepackage[caption=false,font=normalsize,labelfont=sf,textfont=sf]{subfig}
\usepackage{textcomp}
\usepackage{stfloats}
\usepackage{url}
\usepackage{verbatim}
\usepackage{graphicx}
\usepackage{cite}

\usepackage{enumitem}
\usepackage{multirow}
\usepackage{booktabs}
\usepackage{bbding}
\usepackage{bm}

\setitemize[1]{itemsep=0pt,partopsep=0pt,parsep=\parskip,topsep=0pt}

\begin{document}

\title{MB-HGCN: A Hierarchical Graph Convolutional Network for Multi-behavior Recommendation}


\author{

Mingshi Yan,
Zhiyong Cheng,
Jing Sun,
Fuming Sun$^{\dag}$,
and
Yuxin Peng,~\IEEEmembership{Senior Member,~IEEE}


  \thanks{$^{\dag}$ Corresponding Author.}
}

\markboth{Journal of \LaTeX\ Class Files,~Vol.~14, No.~8, June~2023}%
{Shell \MakeLowercase{\textit{et al.}}: A Sample Article Using IEEEtran.cls for IEEE Journals}



\maketitle




\input{00_abstract.tex}

\input{01_introduction.tex}

\input{02_related_work.tex}
\input{03_methodology.tex}
\input{04_experiment.tex}
\input{05_conclusion.tex}

\ifCLASSOPTIONcaptionsoff
  \newpage
\fi
\bibliographystyle{IEEEtran}
\bibliography{introduce}

\end{document}

%% file: 00_abstract.tex
\begin{abstract}

    Collaborative filtering-based recommender systems that rely on a single type of behavior often encounter serious sparsity issues in real-world applications, leading to unsatisfactory performance. 
    Multi-behavior Recommendation (MBR) is a method that seeks to learn user preferences, represented as vector embeddings, from auxiliary information. By leveraging these preferences for target behavior recommendations, MBR addresses the sparsity problem and improves the accuracy of recommendations.
    In this paper, we propose MB-HGCN, a novel multi-behavior recommendation model that uses a hierarchical graph convolutional network to learn user and item embeddings from coarse-grained on the global level to fine-grained on the behavior-specific level.
    Our model learns global embeddings from a unified homogeneous graph constructed by the interactions of all behaviors, which are then used as initialized embeddings for behavior-specific embedding learning in each behavior graph. We also emphasize the distinct of the user and item behavior-specific embeddings and design two simple-yet-effective strategies to aggregate the behavior-specific embeddings for users and items, respectively. Finally, we adopt multi-task learning for optimization. 
    Extensive experimental results on three real-world datasets demonstrate that
    our model significantly outperforms the baselines, achieving a relative improvement of 73.93\% and 74.21\% for \textit{HR@10} and \textit{NDCG@10}, respectively, on the Tmall datasets.
\end{abstract}



%% file: 01_introduction.tex
\section{Introduction}
\IEEEPARstart{P}{ersonalized} recommendation is one of the most effective techniques for addressing the problem of information overload, and it has been widely deployed in various information systems~\cite{zhang2019deep}. Due to its simplicity and effectiveness, Collaborative Filtering (CF)~\cite{Cheng2019mmalfm, KorenBV09, XuZCLS20} has become the mainstream approach in contemporary recommender systems. Over the past few decades, many CF-based models have been developed~\cite{Koren08, NingK11, BPR, KabburNK13}, 
ranging from the early matrix factorization (MF) based methods~\cite{KorenBV09, PMF, BPR}, to deep neural network (DNN) based methods~\cite{NCF, WD}, and more recently, graph neural network (GNN) based methods~\cite{LightGCN, KGAT}. The  rapid advancement of recommendation techniques has greatly enhance the recommendation performance. Since CF-based models mainly rely on the interactions between users and items to learn user preference and make recommendations, an inherent limitation lies in those models is that their performance will degrade sharply when the available interactions are sparse. 

Most existing CF-based methods consider a single behavior, which is usually the target behavior on the platforms  (\textit{e.g.}, \emph{buy} on e-commerce platforms) for modeling. However, such behaviors are often very sparse in real-world systems, leading to a serious sparsity problem in these models.
In reality, users engage in various types of behaviors (\textit{e.g.}, \emph{view} and \emph{collect}) to interact with items and gather information before making a final decision (\textit{i.e.}, engaging in the target behavior).
Those behaviors also contain valuable user preference information, and their interactions are typically richer than those of the target behavior. Therefore, they can be leveraged to learn user preference and alleviate the sparsity problem. 


The utilization of other behaviors (also called auxiliary behaviors) to facilitate the recommendation of target behavior is called multi-behavior recommendation (MBR), which has gained increasing attention in recent years~\cite{NMTR, MBGCN, BPRH, SchlichtkrullKB18, GNMR, ZhangMCX20}. The key to MBR is how to utilize the auxiliary information to assist user and item embedding learning. Earlier approaches are straightforward to extend the traditional matrix factorization techniques operating on single matrix to multiple matrices~\cite{CMF, ZhaoCHC15} or enriched the training data with auxiliary behavior data using different sampling strategies~\cite{LoniPLH16, DingY0QLCJY18, GuoQTLMW17}. With increasing evidence that MBR is effective, it has attracted more attention and recent advanced techniques have aslo been progressively introduced to this task. For instance, NMTR~\cite{NMTR} combines DNNs to model the sequence of behaviors, and MATN~\cite{MATN} adopts the multi-head attention mechanism to model multiple behaviors. Furthermore, GCN-based methods employ various strategies on the unified graph constructed by all behaviors to learn user preferences~\cite{MBGCN, GNMR, MB-GMN, ZhuLLSLCWN23}.  MBGCN~\cite{MBGCN} constructs a heterogeneous graph that distinguishes different types of behaviors with different edges to model each behavior separately, and then aggregates user embeddings by its importance for prediction.  

The basic assumption behind MBR models is that the interaction information of different behaviors contains user preferences from different perspectives or to different extents. A common paradigm of existing DNN- or GNN-based MBR models is to first learn user and item embeddings from each behavior via a designed network and then aggregate the learned embeddings with different strategies for target behavior prediction (\textit{e.g.}, with or without attention mechanism)~\cite{MATN, MBGCN, GNMR, MB-GMN, SMBREC, DIPN}. The difference lies in how to design the network structure to learn better embeddings from different behaviors and how to distill valuable information from each behavior to contribute to the target behavior prediction. 
 
In this work, we propose a novel MBR model with a hierarchical graph convolutional network (MB-HGCN) to utilize auxiliary behaviors for user and item embedding learning. Unlike previous GCN-based methods that directly learn user and item embeddings from the unified heterogeneous graph, our model adopts a different paradigm to learn user and item embeddings via a hierarchical network structure. Specifically, we first learn a global embedding for each user and item in a unified homogeneous graph constructed based on the interactions of all behaviors without differentiating the behavior types. We then take the global embeddings as initialized embeddings to each behavior-specific graph, which is constructed based on the interactions of each behavior, for subsequent behavior-specific embedding learning. By performing graph convolutional operations on the unified homogeneous graph, which contains all the interaction information of different behaviors, we can fully exploit all the interaction information to learn the global embeddings of the users and items. Although the global embeddings might be of coarse-grained as they have not differentiated different behavior types, they could be a good initial embeddings for the following behavior-specific embedding learning on each behavior graph. This can also alleviate the sparsity issue in each behavior, as a good embedding initialization is crucial for the representation learning in deep models. The behavior-specific embedding learning on each individual behavior graph aims to capture behavior-specific features for better preference learning. 
 
After the two-stage of embedding learning, we aggregate the behavior-specific embeddings with two different strategies for user and item embedding aggregations, respectively. For user embedding, to distill effective information from different behaviors for the target behavior prediction, we assign the weight to a behavior-specific embedding based on its similarity to the target behavior-specific embedding, with the intuition that the more similar the two embeddings are, the more they contribute to the target behavior. 
For item embeddings, we adopt a weighting scheme based on the interaction numbers of different behaviors~\cite{MBGCN}. The rationale behind this design is that item features should be consistent across different behaviors, and the difference between item embeddings learned from different behaviors is caused by the interactions of different users.
Finally, we combine the global embedding and the aggregated behavior-specific embeddings for more comprehensive representations. Multi-task learning is adopted to treat each behavior as an independent task in optimization. To evaluate the effectiveness of our model, we perform extensive empirical studies on three large-scale real-world datasets. The experimental results demonstrate that our model outperforms the state-of-the-art MBR models by a large margin. For example, on the Tmall dataset, our model achieve an impressive improvement of 73.93\% and 74.21\% over the second-best baseline in terms of HR@10 and NDCG@10, respectively. We also conduct comprehensive ablation studies to carefully examine the utility of different designs in our model.

In summary, the main contributions of this work are as follows:
\begin{itemize}[leftmargin=*]
    \item We propose a hierarchical convolutional graph network for multi-behavior recommendation that learns user and item embeddings from the coarse-grained and global level with a unified graph to the fine-grained and behavior-specific level in each behavior graph. 
    We deem this learning paradigm can better utilize the multi-behavior information to learn good user and item embeddings.
    
    
    \item We emphasize the distinctiveness of the user and item behavior-specific embeddings, for which we design two simple-yet-effective aggregation strategies to aggregate the behavior-specific embeddings for users and items, respectively. This is quite different from mainstream aggregation methods that use the same mechanism for aggregation.
    
    \item We conduct extensive experiments on three real-world datasets to evaluate the effectiveness of our MB-HGCN model and examine the validity of each component in MB-HGCN. Experimental results show that MB-HGCN achieves a remarkable improvement over the state-of-the-art models in terms of recommendation accuracy. Additionally, we release the codes and involve parameters to benefit other researchers~\footnote
{https://github.com/MingshiYan/MB-HGCN.}. 
  \end{itemize}
  
The rest of this paper is structured as follows. Section~\ref{Related Work} reviews the related work, and Section~\ref{methodology} describes our MB-HGCN model in detail. Next, Section~\ref{experiment} introduces the experimental setup and reports the experimental results. Finally, Section~\ref{conclusion} concludes this paper.

%% file: 02_related_work.tex
\section{Related Work} \label{Related Work}
Multi-behavior recommendation refers to leveraging user-item interaction data of multi-type behaviors for recommendation~\cite{NMTR, DingY0QLCJY18, MBGCN}. Its advantage is to alleviate the data sparsity problem existing in single-behavior-based recommendation methods. Due to its excellent performance, it has attracted increasing attention in recent years~\cite{CKML, BPR, GNMR}.

Early multi-behavior recommendation methods are extended on traditional CF-based methods~\cite{CMF, ZhaoCHC15, LoniPLH16, DingY0QLCJY18, GuoQTLMW17, BPRH}, and the most direct approach is to apply the matrix factorization approach in single-behavior data into multi-behavior data. For example, Ajit et al.~\cite{CMF} proposed a collective matrix factorization model (CMF), in which entity parameters are shared among multiple matrix factorizations. This method is extended by Zhao et al.~\cite{ZhaoCHC15} to perform matrix factorization for different behaviors by sharing item embeddings. In addition, some researchers designed different sampling strategies to utilize the data from multiple behaviors. For example, Loni et al.~\cite{LoniPLH16} proposed a negative sampling strategy suitable for multiple behaviors to sample user-item interaction data of different behaviors. Ding et al.~\cite{DingY0QLCJY18} put forward an improved negative sampling strategy to achieve better utilization of data, further extending this idea. Guo et al.~\cite{GuoQTLMW17} introduced a strategy of sampling based on similarity, which generates positive and negative samples from multiple auxiliary behaviors to help model training. 
Qiu et al.~\cite{BPRH} proposed an adaptive sampling strategy according to the uncorrelated balance characteristics of samples between different behaviors.
These methods supplement the training on the target behavior by exploiting the interaction data from the auxiliary behaviors. 


With the development of deep learning, multi-behavior recommendation methods based on deep neural networks (DNN) have been developed~\cite{MATN, DIPN, NMTR}. The main idea of such models is to design deep neural networks to learn the embedding of users and items separately from each behavior, and then aggregate them for recommendation. The difference between these methods is mainly reflected in the design of DNN and the aggregation strategies. For example, Xia et al.~\cite{MATN} designed a network consisting of transformers and multi-head attention mechanisms to learn embeddings in each behavior, and then aggregated them by adopting a fully connected network. Guo et al.~\cite{DIPN} proposed a hierarchical attention mechanism to aggregate user preferences learned from different behaviors. Unlike other methods that aggregate information learned from different behaviors for prediction, Gao et al.~\cite{NMTR} adopted a sequential modeling approach to explore the dependencies of different behaviors by passing the current behavior prediction score forward. The advantage of deep networks in representation learning makes the DNN-based MBR models achieve great progress on recommendation performance.

Recently,  with the success of  graph convolutional networks in recommendation, many  GCN-based MBR methods have also been proposed~\cite{MB-GMN, MBGCN, SMBREC, GHCF, CKML, GNMR, CRGCN}. Similar to DNN-based models, the general paradigm for such methods is to model each behavior separately with GCN to learn the embeddings of users and items, and then aggregate them with different strategies. For example, Xia et al.~\cite{MB-GMN} proposed a multi-behavior pattern encoding framework and a graph element network to explore complex dependencies between different types of user-item interactions. Jin et al.~\cite{MBGCN} performed user-item propagation and item-item propagation in different behaviors on a multi-behavior heterogeneous graph to learn the influence strength and semantics of different behaviors. Chang et al.~\cite{CKML} proposed a multi-interest learning framework, containing an interest-extracting module and a behavioral correlation module, to better model the complex dependencies among multiple behaviors. 
Gu et al.~\cite{SMBREC} designed different strategies to aggregate the embeddings of multi-behavior users and items separately, and adopted a star-shaped contrastive learning to capture the commonality between target behaviors and auxiliary behaviors. Different from those GCN-based methods, Yan et al.~\cite{CRGCN} proposed a cascaded residual network to explore the connection between different behaviors from the perspective of embedding propagation. 

In this work, we propose a novel hierarchical GCN structure to exploit the multi-behavior data for user and item embedding learning with a different paradigm. Our model first learn the global embeddings from a unified homogeneous graph constructed on all the behavior data, and then take them as initial embeddings for subsequent behavior-specific embedding learning. This learning strategy can well utilize the multi-behavior information and promise a good embedding initialization for the embedding learning in each behavior graph. Besides, we adopt two different strategies to aggregate the behavior-specific embeddings for users and items. The comparisons with existing advanced MBR models in empirical study demonstrates the effectiveness of our model.



%% file: 03_methodology.tex
\section{methodology} \label{methodology}

\begin{figure*}[hbt]
    \centering
    \includegraphics[width=1.\linewidth]{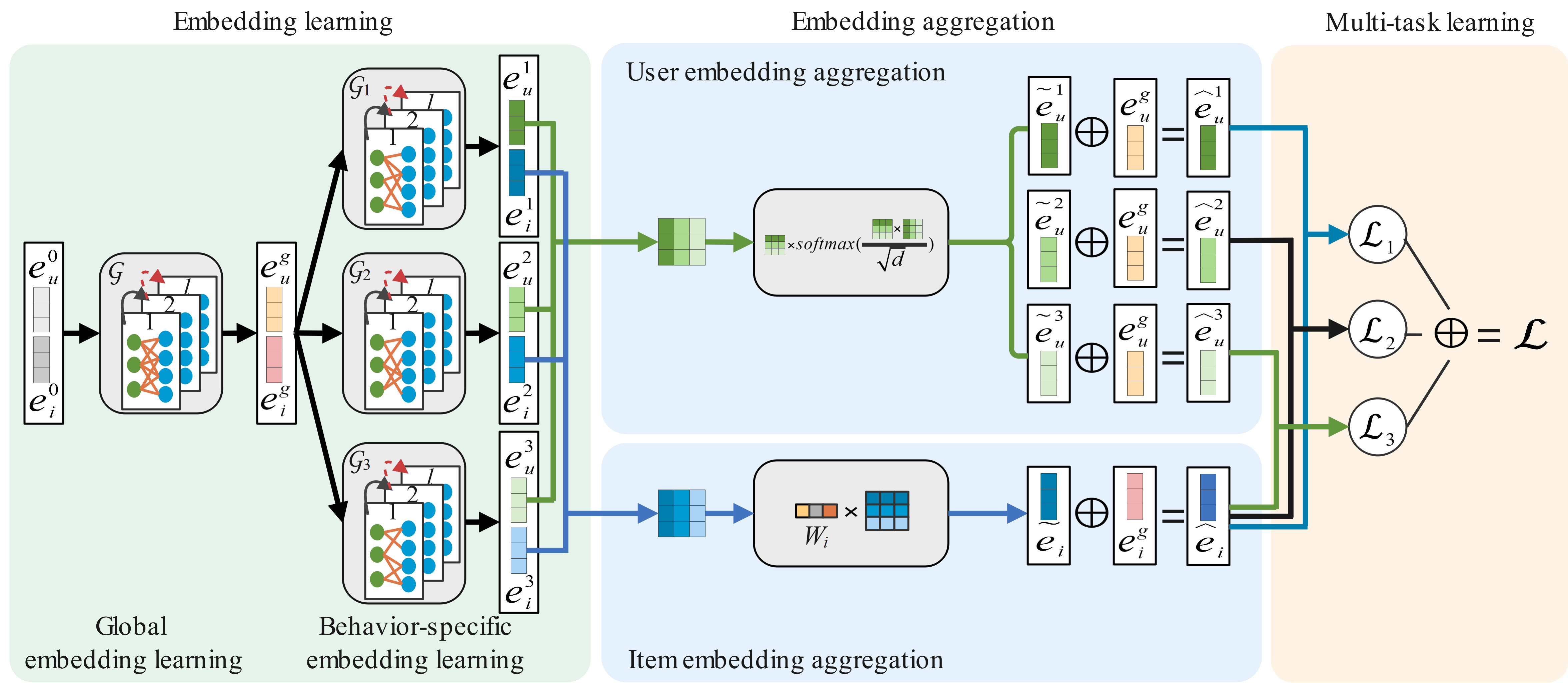}
    \caption{Overview of MB-HGCN model. We take an example that \textit{K=3}, \textit{i.e.}, there are three kinds of behavior, in which the
    last is target behavior and the other two are auxiliary behaviors.}
    \label{fig:global}
  
\end{figure*}

\subsection{Preliminaries}
Multi-behavior recommendation (MBR) is to utilize auxiliary behaviors (\textit{e.g.}, \emph{view} and \emph{cart}) when interacting with the platforms to help learn user preferences. These behaviors also reflect users' interests in items and thus contains rich user preference information, which can be leveraged to effectively alleviate the data sparsity problem. In this work, we aim to learn better user and item embeddings by exploiting auxiliary behaviors to improve the recommendation performance.

Let $\mathcal{U}$ and $\mathcal{I}$ be the set of users and items, and the total number of users and items are $M$ and $N$, respectively.  $K$ is the number of behavior types. We use $k (1 \leq k \leq K)$ to represent the $k$-th behavior, and the $K$-th behavior is the target behavior. Let $\mathcal{R}_k$ be the interaction matrix for the $k$-th behavior, which is a binary matrix. For $r_{ui} \in \mathcal{R}_k$, $r_{ui}=1$ if there was an interaction of the $k$-th behavior happened between user $u$ and item $i$; otherwise $r_{ui}=0$.

The studied problem is formulated as follows:

\textbf{Input:} user set $\mathcal{U}$, item set $\mathcal{I}$,  and user-item interaction matrices $\{\mathcal{R}_1, \cdots, \mathcal{R}_K\}$ for different types of behaviors.

\textbf{Output:} a similarity score, which indicates the possibility that a user $u$ will interact with an item $i$ in the target behavior.

Before describing our model, we would like to first introduce  two types of graphs used in our model:

\begin{itemize}[leftmargin=*]
\item \textbf{Behavior-specific graph}, denoted by  $\mathcal{G}_k=(V_k, E_k)$, which is a bipartite graph constructed based on the interactions of the $k$-th behavior type according to the interaction matrix $\mathcal{R}_k$. $V_k$ consists of the user node $u \in \mathcal{U}$ and the item node $i \in \mathcal{I}$, and $E_k$ denotes the user-item interaction edges in graph $\mathcal{G}_k$. There is an edge between a user node and an item node if $r_{ui}=1$ for $ r_{ui} \in \mathcal{R}_k$.

\item \textbf{Unified graph}, denoted by $\mathcal{G}=(V, E)$, which is constructed based on the interactions of all types of behaviors. It is a homogeneous graph, which means we do not differentiate the different types of interactions in this graph. For interactions of different types or multiple interactions of different behaviors between a user $u$ and an item $i$, the edge is the same, namely, $E=E_1 \cup E_2 \cup \cdots \cup E_K$.   
\end{itemize}



\subsection{Model Description}

\textbf{Overview.} Users' interaction behaviors with items reflect their interests. In multi-behavior recommendation, it is well-recognized that different types of behaviors disclose user's preference from different perspectives or to different extents~\cite{CKML, WanM18, NMTR}. Based on this common assumption, many MBR approaches have been proposed to extract valuable information from multiple behaviors to learn user preferences. Most previous MBR models first learn embeddings from different behaviors separately and then aggregate them with different strategies. The ultimate goal is to exploit the auxiliary behaviors to learn better user and item embeddings, thereby enhancing the recommendation accuracy of the target behavior. 

In this work, we propose a hierarchical graph convolutional network to exploit the multi-behavior information to learn the user and item embeddings. In particular, we first learn a global embedding by adopting the unified graph constructed based on interaction information of all behaviors. The global embedding is then used as the initialized embedding and fed into the behavior-specific graph to learn behavior-specific embeddings for each type of behavior. The intuition is that there is a general interest of users across different behaviors and each behavior contains some distinct features of user preference. The global embedding learned from the unified graph represents the general interests or coarse-grained preferences, and the behavior-specific embedding learned from each behavior-specific graph represents the refined or fine-grained user preferences for this particular behavior. In the next, two different strategies will be used to respectively obtain the final user and item embeddings for prediction. Multi-task learning is used for optimization. 


Fig.~\ref{fig:global} shows the overall structure of our MB-HGCN model, which mainly consists of three modules: 1) \textbf{Embedding learning}, which is designed to learn the embeddings of users and items via a hierarchical graph network structure; 2) \textbf{Embedding aggregation}, which adopts two different strategies to aggregate the embeddings of users and items. More specific, a novel weighting scheme is designed to adaptive distill valuable information from different behaviors for user embedding aggregation; and a linear aggregation approach is used for item embedding aggregation; 3) \textbf{Multi-task learning} is adopted to employ the interaction information of each behavior as supervision signals for user and item embeddings. In the following subsections, we will first brief the embedding initialization and then describe the three modules detailedly in sequence.

\subsubsection{\textbf{Embedding initialization}}
Following previous works~\cite{NMTR, LightGCN, MBGCN, UltraGCN}, we initialize the ID of user $u \in \mathcal{U}$ and item $i \in \mathcal{I}$ as $d$-dimensional embedding vector $\bm{e_u}^0$ and $\bm{e_i}^0$, respectively. Let $\boldsymbol{P} \in \mathbb{R}^{M \times d}$ and $\boldsymbol{Q} \in \mathbb{R}^{N \times d}$ be the embedding matrices for the user and item embedding initialization, where  $M$ and $N$ represent the number of users and items, respectively. Each user and item ID is represented as a unique embedding. Given the one-hot embedding matrix  $\boldsymbol{ID}^{\mathcal{U}}$ and $\boldsymbol{ID}^{\mathcal{I}}$ for all users and items,  the embeddings of user $u$ and item $i$ are initialized as:

\begin{equation}
  \label{eq:init}
  \begin{aligned}
    \boldsymbol{e}_{u}^0 = \boldsymbol {P} \cdot \boldsymbol{ID}_u^{\mathcal{U}}, \quad
    \boldsymbol{e}_{i}^0 = \boldsymbol {Q} \cdot \boldsymbol{ID}_i^{\mathcal{I}},
  \end{aligned}
\end{equation}
where $\boldsymbol{ID}_u^{\mathcal{U}}$ and $\boldsymbol{ID}_i^{\mathcal{I}}$ represent the user $u$'s and the item $i$'s one-hot vector, respectively.

\subsubsection{\textbf{Embedding Learning}}
As mentioned, our model adopts a hierarchical GCN structure to exploit the multi-behavior for embedding learning. The interaction information of all behaviors are integrated into the unified graph $\mathcal{G}$ to learn the general user preferences and item features, denoted as the global embedding $\bm{e}_u^{g}$ and $\bm{e}_i^{g}$ for user $u$ and item $i$, respectively. The learned global embedding is then fed into each behavior-specific graph $\mathcal{G}_k$ to learn the behavior-specific embedding $\bm{e}_u^{k}$ and $\bm{e}_i^{k}$ for user $u$ and item $i$ in $\mathcal{G}_k$, respectively. For the embedding learning in each graph, we employ the LightGCN~\cite{LightGCN} model, which is a lightweight CF-based single-behavior recommendation model. It simplifies the standard GCN and only retains the core neighborhood aggregation component, which has proven to be effective and superior in performance.  It is worth mentioning that other GCN  models can also be adopted, such as Ultra-GCN~\cite{UltraGCN} and SVD-GCN~\cite{SVD-GCN}. 
The core of LightGCN is to recursively aggregate information from neighboring nodes for embedding update of the target node. The graph convolution operation in LightGCN is:
\begin{equation}
    \label{eq:agg}
    \begin{aligned}
      \boldsymbol e_{u}^{(l+1)} &= \sum_{i \in N_{u}} \frac{1}
      {\sqrt{\left\lvert N_{u} \right\rvert} \sqrt{\left\lvert N_{i} \right\rvert}} \boldsymbol e_{i}^{(l)}, \\ 
      \boldsymbol e_{i}^{(l+1)} &= \sum_{u \in N_{i}} \frac{1}
      {\sqrt{\left\lvert N_{i} \right\rvert} \sqrt{\left\lvert N_{u} \right\rvert}} \boldsymbol e_{u}^{(l)}, 
    \end{aligned}
\end{equation}
where  $\frac{1}{\sqrt{\left\lvert N_{u} \right\rvert} \sqrt{\left\lvert N_{i} \right\rvert}}$ denotes the normalization coefficient, $N_{u}$ represents the set of items that are interacted with the user $u$, and $N_{i}$ is the same. After \textit{l}-layer propagation, LightGCN combines the embeddings obtained at each layer as the final user and item representation. Given the total number of layers as $L$, the representation of user $u$ and item $i$ after the LightGCN process are as follows:
\begin{equation}
  \label{eq:comb}
  \begin{aligned}
    \boldsymbol{e}'_{u} = \sum_{l=0}^{L}{\alpha_l \boldsymbol{e}_u^{(l)}}, \quad
    \boldsymbol{e}'_{i} = \sum_{l=0}^{L}{\alpha_l \boldsymbol{e}_i^{(l)}},
  \end{aligned}
\end{equation}
where $\alpha_l$ is a hyperparameter represents the importance of the $l$-th layer embedding, and $\bm{e}_u^{(0)} (\boldsymbol{e}_i^{(0)})$ is the initial embedding of user $u$ (item $i$). 


As shown in Fig.~\ref{fig:global}, LightGCN with the same settings is used in both the unified graph $\mathcal{G}$ and each behavior-specific graph $\mathcal{G}_k$ to learn the embeddings of users and items. 

\textbf{Global embedding.} Following the Eq.~\ref{eq:agg}, we respectively obtain \textit{L} embeddings to describe a user $\{\boldsymbol{e}_u^{(1)}, \boldsymbol{e}_u^{(2)}, \cdots, \boldsymbol{e}_u^{(L)} \}$ and an item $\{\boldsymbol{e}_i^{(1)}, \boldsymbol{e}_i^{(2)}, \cdots, \boldsymbol{e}_i^{(L)} \}$ after \textit{L-layer} propagation. Before combining these embeddings, normalization is adopted to alleviate the impact of the embedding scale~\cite{LightGCN}. In our model, we apply $L_2$ normalization for simplicity:
\begin{equation}
  \label{eq:comb0}
  \begin{aligned}
    \boldsymbol{e}^{g}_{u} = \boldsymbol{e}_{u}^{(0)} + \sum_{l=1}^{L}{\alpha_l \frac{\boldsymbol{e}_u^{(l)}}{\lVert \boldsymbol{e}_u^{(l)} \rVert_{2}}}, \quad
    \boldsymbol{e}^{g}_{i} = \boldsymbol{e}_{i}^{(0)} + \sum_{l=1}^{L}{\alpha_l \frac{\boldsymbol{e}_i^{(l)}}{\lVert \boldsymbol{e}_i^{(l)} \rVert_{2}}},
  \end{aligned}
\end{equation}
where $\boldsymbol{e}^{g}_{u}$ and $\boldsymbol{e}^{g}_{i}$ are the  learned global embedding from $\mathcal{G}$, they are also the sharing input of the following behavior-specific graph $\mathcal{G}_k$. $\bm{e}_{u}^{(0)}$ and $\bm{e}_{i}^{(0)}$ are the input of graph $\mathcal{G}$ (\textit{i.e.}, the initialized embedding $\bm{e}_{u}^0$ and $\bm{e}_{i}^0$). Intuitively, the farther neighbors have less important, thus, we set the $\alpha_l$ as $1/(l + 1)$. 

\textbf{Behavior-specific embedding.} Similarly, taking $\boldsymbol{e}^{g}_{u}$ and $\boldsymbol{e}^{g}_{i}$ as the initial embeddings for the embedding learning in each behavior-specific graph $\mathcal{G}_k$, we can obtain $K$ behavior-specific embeddings for each user $u$ and item $i$ ( \textit{i.e.}, $\{\boldsymbol{e}_u^{1}, \boldsymbol{e}_u^{2}, \cdots, \boldsymbol{e}_u^{K}\}$ and $\{\boldsymbol{e}_i^{1}, \boldsymbol{e}_i^{2}, \cdots, \boldsymbol{e}_i^{K}\}$).

\subsubsection{\textbf{Embedding Aggregation}} 
Through the embedding learning process, for each user $u$ and item $i$,  we obtain their global embeddings $\boldsymbol{e}^{g}_{u}$ and $\boldsymbol{e}^{g}_{i}$, as well as behavior-specific embedding set $\{\boldsymbol{e}_u^{1}, \boldsymbol{e}_u^{2}, \cdots, \boldsymbol{e}_u^{K}\}$  and $\{\boldsymbol{e}_i^{1}, \boldsymbol{e}_i^{2}, \cdots, \boldsymbol{e}_i^{K}\}$. In the next, we would like to aggregate the above user and item embeddings respectively by adopting different strategies to obtain the final embedding for recommendation. 

\textbf{User embedding aggregation.} Considering different behaviors may convey some distinct information of user preference, to distill valuable information from different behavior-specific embeddings for the target behavior prediction, we design a novel weighting scheme for user embedding aggregation. Taking the user $u$ as example, the aggregation is formulated as:
\begin{equation}
  \label{eq:concat1}
  \begin{aligned}
    \boldsymbol{U} = \boldsymbol{e}_u^{1} || \boldsymbol{e}_u^{2} || \cdots || \boldsymbol{e}_u^{K}, 
  \end{aligned}
\end{equation}
\begin{equation}
  \label{eq:att_user}
  \begin{aligned}
    \tilde{\boldsymbol{e}}_u^{k} = \boldsymbol{U} \boldsymbol{\delta}^{\top},
  \end{aligned}
\end{equation}
where $||$ denotes the operation of stacking vectors to form a matrix. $\boldsymbol{U} \in \mathbb{R}^{d \times K}$ is the matrix by stacking user embeddings learned from each behavior, in which $d$ represents the embedding size and $K$ is the number of behaviors. $\boldsymbol{\delta}$ is the weight vector calculated based on the similarity between the embedding of each behavior and that of the target behavior. Formally, it is computed as

\begin{equation}
  \label{eq:w_delta}
  \begin{aligned}
    \boldsymbol{\delta} = softmax(\frac{{\boldsymbol{e}_u^k}^{\top} \boldsymbol{U}}{\sqrt{d}}).
  \end{aligned}
\end{equation}


In this equation, ${\boldsymbol{e}_u^k}^{\top} \boldsymbol{U}$ calculates the embedding similarity between the $k$-th behavior and other behaviors. The denominator $\sqrt{d}$ is used to prevent the vanishing gradient problem, and $softmax(\cdot)$ is adopted for normalization. 

The underlying rationality of the aggregation strategy is that the behavior with more similar embeddings contain more relevant preference information with the target behavior, and thus contribute more to the target behavior in the aggregation.  With this aggregation strategy, our model can adaptively extract valuable information from other behaviors for the target behavior prediction. Moreover, together with the multi-task learning, it also avoids the behavior-specific embeddings to be optimized towards the target behavior in the learning process. 
Through the above operation, we can obtain the embedding set $\{\tilde{\boldsymbol{e}}_u^{1}, \tilde{\boldsymbol{e}}_u^{2}, \cdots, \tilde{\boldsymbol{e}}_u^{K}\}$.


\textbf{Item embedding aggregation.} Since item features are consistent across different behaviors, we simply apply a linear combination  to aggregate the embeddings learned from different behaviors. In different types of behaviors, the users who interacted with the items are different and the total number of interactions (\textit{i.e.}, the number of users who interacted with the items) are also different. Intuitively, with more interactions, the learned features should be more comprehensive. Accordingly, the weight assigned to the $k$-th behavior-specific embedding for an item $i$ is defined as:
\begin{equation}
  \label{eq:weight}
  \begin{aligned}
    \gamma_{ik} = \frac{w_k \cdot n_{ik}}{\sum_{m=1}^{K}{w_m \cdot n_{im}}},
  \end{aligned}
\end{equation}
where $w_k$ is a learnable parameter for the $k$-th behavior; $n_{ik}$ denotes the number of users that interacted with item $i$ in the $k$-th behavior. The behavior-specific embeddings of item $i$ are aggregated as:
\begin{equation}
  \label{eq:att_item}
  \begin{aligned}
    \tilde{\boldsymbol{e}}_i = \sum_{k=1}^{K}{\gamma_{ik} \cdot \boldsymbol{e}_i^{k}}.
  \end{aligned}
\end{equation}

Notice that both user embedding $\tilde{\boldsymbol{e}}_u^{k}$ and item embedding $\tilde{\boldsymbol{e}}_i$ are obtained from the behavior-specific information. In order to obtain more comprehensive representation, we combine them with the global embeddings to obtain the final embeddings:
\begin{equation}
  \label{eq:fuse_u}
  \begin{aligned}
    \hat{\boldsymbol{e}}_u^{k} = \tilde{\boldsymbol{e}}_u^{k} \oplus \boldsymbol{e}_u^{g}, \quad
    \hat{\boldsymbol{e}}_i = \tilde{\boldsymbol{e}}_i \oplus \boldsymbol{e}_i^{g},
  \end{aligned}
\end{equation}
where $\oplus$ denotes the element-wise sum.

\subsubsection{\textbf{Multi-task Learning (MTL)}}  MTL~\cite{PLE} is a learning strategy for jointly optimizing different-yet-related tasks. To better exploit the multiple-behavior information in user and item embedding learning, we treat each behavior as an independent training task. 
The inner product of user and item embeddings is adopted to estimate the prediction score. Take the $k$-th behavior as an example:
\begin{equation}
  \label{eq:score}
  y_{ui}^{k} = \hat{\bm{e}}_{u}^{k^{\top}} \hat{\bm{e}}_{i}.
\end{equation}

The pairwise Bayesian Personalized Ranking (BPR)~\cite{BPR} loss is adopted in optimization for each task:
\begin{equation}
  \label{eq:loss_func}
  \mathcal{L}_k = \sum_{(u,i,j) \in \mathcal{O}} -ln \sigma(y_{ui}^{k}-y_{uj}^{k}),
\end{equation}
where $\mathcal{O}=\{(u,i,j)|(u,i) \in \mathcal{R}^{+}, (u,j) \in \mathcal{R}^{-}\}$ is defined as positive and negative sample pairs, $\mathcal{R}^{+}$ ($\mathcal{R}^{-}$) denotes the observed (unobserved) samples in the $k$-th behavior, and $\sigma(\cdot)$ is the sigmoid function. Following the Eq.~\ref{eq:loss_func}, we obtain the loss function for all the $K$ tasks, \textit{i.e.}, $\{\mathcal{L}_1, \mathcal{L}_2, \cdots, \mathcal{L}_K\}$, then the $K$ loss functions are summed for joint optimization. Intuitively, the contribution of different tasks should be different. Assigning different weights to different losses may enhance the final performance, however, this is not our main focus in this study. Here we simply treat them equally to focus on studying the effectiveness of our embedding learning strategy  and leave the study of different weights in the loss function as a future work. The final loss function is formulated as:
\begin{equation}
  \label{eq:total_loss}
  \mathcal{L} = \sum_{k=1}^{K} \mathcal{L}_k + \beta \cdot \left\lVert \boldsymbol \Theta \right\rVert_{2},
\end{equation} 
where $\boldsymbol \Theta$ represents all trainable parameters in our model and $\beta$ is the coefficient that controls the strength of the $L_2$ normalization to prevent over-fitting. To improve the generalization ability, two widely used dropout strategies~\cite{MBGCN, NGCF, GCMC} are also adopted in training: node dropout and message dropout, which are used to randomly drop out nodes in the graph and information in the embedding, respectively.

%% file: 04_experiment.tex
\section{experiment} \label{experiment}
In this section, we conduct extensive experiments on three real-word datasets to evaluate the effectiveness of our model. In particular, we aim to answer the following research questions:
\begin{itemize}[leftmargin=*]
    \item \textbf{RQ1:} How does our MB-HGCN model perform as compared with the state-of-the-art recommendation models that are learned from single- and multi-behavior data?
    \item \textbf{RQ2:} How does the key designs in our MB-HGCN model affect the recommendation performance?
    \item \textbf{RQ3:} How does the layer numbers of GCN setting in the LightGCN affect the performance of our model?
    \item \textbf{RQ4:} Can MB-HGCN alleviate the cold start problem?
    \item \textbf{RQ5:} How does the user embedding learned in the MB-HGCN model?
\end{itemize}

\subsection{Experiment Settings} \label{Experiment Settings}

\subsubsection{\textbf{Dataset}} Three real-world datasets are adopted for experiments:

\begin{itemize}[leftmargin=*]
  \item \textbf{Tmall.} This dataset is collected from Tmall\footnote{https://www.tmall.com/}, which is one of the largest e-commerce platforms in China. It contains 41,738 users and 11,953 items with 4 types of behaviors, \textit{i.e.}, \emph{view}, \emph{collect}, \emph{cart}, and \emph{buy}. 

  \item \textbf{Beibei.} This dataset is collected from Beibei\footnote{https://www.beibei.com/}, which is the largest infant product retail e-commerce platform in China. This dataset contains 21,716 users and 7,977 items with three types of behaviors, \textit{i.e.}, \emph{view}, \emph{cart}, and \emph{buy}.
  
  \item \textbf{Jdata.} This dataset is collected from JD\footnote{https://www.jd.com/}, which is one of the most popular and influential e-commerce websites in the Chinese e-commerce field. This dataset contains 93,334 users and 24,624 items with 4 types of behaviors, \textit{i.e.}, \emph{view}, \emph{collect}, \emph{cart}, and \emph{buy}.

\end{itemize} 

For the above datasets, we follow the previous work to remove the duplicated records by keeping the earliest one~\cite{NMTR, MBGCN}. The statistical information of the three datasets is summarized in Table~\ref{tab:dataset}.

\begin{table}[htb]
  \caption{Statistics of three real-world benchmark datasets.}
  \label{tab:dataset}
  \resizebox{\columnwidth}{!}{
    \begin{tabular}{ccccccc}
      \toprule
      \textbf{Dataset} & \textbf{Users} & \textbf{Items} & \textbf{Buy} & \textbf{Cart} & \textbf{Collect} & \textbf{View} \\
      \midrule
      \textbf{Tmall}  & 41,738 & 11,953 & 255,586 & 1,996   & 221,514 & 1,813,498 \\
      \textbf{Beibei} & 21,716 & 7,997  & 304,576 & 642,622 & -       & 2,412,586 \\
      \textbf{Jdata}  & 93,334 & 24,624 & 333,383 & 49,891  & 45,613  & 1,681,430 \\
      \bottomrule
    \end{tabular}
  }
\end{table}

\subsubsection{\textbf{Evaluation Protocols}}
We adopt the widely used leave-one-out strategy for model evaluation~\cite{Gao0GCFLCJ19, MBGCN, NMTR}. In training stage, the last postive item for each user is selected to construct the validation set for hyper-parameter tuning. In the evaluation stage, all the items in the test set are ranked according to the predicted scores by recommendation models. Meanwhile, two representative evaluation metrics in recommendation: \emph{Hit Ratio (HR@K)}~\cite{HR} and \emph{Normalized Discounted Cumulative Gain (NDCG@K)}~\cite{NDCG} are adopted to evaluate the performance:
\begin{itemize}[leftmargin=*]
  \item \textbf{HR@K:} a performance metric used to evaluate the accuracy of a recommender system by measuring the proportion of test items for which the correct recommendation appears within the top K positions of the ranked list.
  \item \textbf{NDCG@K:} a metric that measures the quality of the recommended items by considering both their relevance and their position in the ranked list.
\end{itemize}

\begin{table*}[hbt]
  \caption{Overall performance comparison (\textit{Impr.} means the relative improvement over the best baselines, optimal and suboptimal are bolded and underlined, respectively).}
  \label{tab:overall}
  \resizebox{\textwidth}{!}{
	\begin{tabular}{crcccccccccccr}
	\toprule
	\multirow{2}{*}{\textbf{Dataset}} & \multirow{2}{*}{\textbf{Metric}} & \multicolumn{3}{c}{\textbf{Single-behavior}}       & \multicolumn{7}{c}{\textbf{Multi-behavior}}                                                                           & \multirow{2}{*}{\textit{\textbf{Impr.}}} \\ \cmidrule(lr){3-5} \cmidrule(lr){6-12}
									                  &                                  & \textbf{MF-BPR} & \textbf{NCF} & \textbf{LightGCN} & \textbf{R-GCN} & \textbf{NMTR} & \textbf{MBGCN} & \textbf{GNMR} & \textbf{S-MBRec} & \textbf{CRGCN} & \textbf{MB-HGCN}   &                               \\ \hline  
	\multirow{6}{*}{\textbf{Tmall}}   & \textbf{HR@10}                   & 0.0230          & 0.0301       & 0.0393            & 0.0316         & 0.0517        & 0.0549         & 0.0393        & 0.0694           & \underline{0.0840}         & \textbf{0.1461} & 73.93\%  \\ 
									                  & \textbf{NDCG@10}                 & 0.0124          & 0.0153       & 0.0209            & 0.0157         & 0.0250        & 0.0285         & 0.0193        & 0.0362           & \underline{0.0442}         & \textbf{0.0770} & 74.21\%  \\ 
									                  & \textbf{HR@20}                   & 0.0316          & 0.0420       & 0.0538            & 0.0489         & 0.0847        & 0.0799         & 0.0619        & 0.1009           & \underline{0.1238}         & \textbf{0.2072} & 67.37\%  \\ 
									                  & \textbf{NDCG@20}                 & 0.0144          & 0.0182       & 0.0243            & 0.0198         & 0.0330        & 0.0345         & 0.0247        & 0.0438           & \underline{0.0540}         & \textbf{0.0920} & 70.37\%  \\ 
									                  & \textbf{HR@50}                   & 0.0434          & 0.0678       & 0.0813            & 0.0826         & 0.1498        & 0.1285         & 0.1071        & 0.1553           & \underline{0.1994}         & \textbf{0.3149} & 57.92\%  \\ 
									                  & \textbf{NDCG@50}                 & 0.0166          & 0.0231       & 0.0295            & 0.0262         & 0.0456        & 0.0438         & 0.0332        & 0.0544           & \underline{0.0685}         & \textbf{0.1130} & 64.96\%  \\ \hline
	\multirow{6}{*}{\textbf{Beibei}}  & \textbf{HR@10}                   & 0.0268          & 0.0296       & 0.0309            & 0.0327         & 0.0315        & 0.0373         & 0.0396        & 0.0489           & \underline{0.0539}         & \textbf{0.0619} & 14.84\%  \\ 
									                  & \textbf{NDCG@10}                 & 0.0139          & 0.0146       & 0.0161            & 0.0161         & 0.0146        & 0.0193         & 0.0219        & 0.0253           & \underline{0.0259}         & \textbf{0.0297} & 14.67\%  \\ 
									                  & \textbf{HR@20}                   & 0.0427          & 0.0453       & 0.0478            & 0.0561         & 0.0587        & 0.0639         & 0.0640        & 0.0770           & \underline{0.0944}         & \textbf{0.1019} &  7.94\%  \\ 
									                  & \textbf{NDCG@20}                 & 0.0179          & 0.0185       & 0.0204            & 0.0219         & 0.0214        & 0.0259         & 0.0280        & 0.0324           & \underline{0.0361}         & \textbf{0.0397} &  9.97\%  \\ 
									                  & \textbf{HR@50}                   & 0.0793          & 0.0809       & 0.0880            & 0.1118         & 0.1276        & 0.1287         & 0.1219        & 0.1234           & \underline{0.1817}         & \textbf{0.2009} & 10.57\%  \\ 
									                  & \textbf{NDCG@50}                 & 0.0250          & 0.0216       & 0.0282            & 0.0329         & 0.0348        & 0.0386         & 0.0394        & 0.0415           & \underline{0.0532}         & \textbf{0.0592} & 11.28\%  \\ \hline
	\multirow{6}{*}{\textbf{Jdata}}   & \textbf{HR@10}                   & 0.1850          & 0.2090       & 0.2252            & 0.2406         & 0.3142        & 0.2803         & 0.3068        & 0.4125           & \underline{0.5001}         & \textbf{0.5338} &  6.74\%  \\ 
									                  & \textbf{NDCG@10}                 & 0.1238          & 0.1410       & 0.1436            & 0.1444         & 0.1717        & 0.1572         & 0.1581        & 0.2779           & \underline{0.2914}         & \textbf{0.3238} & 11.12\%  \\ 
									                  & \textbf{HR@20}                   & 0.2192          & 0.2461       & 0.2825            & 0.3418         & 0.4086        & 0.3603         & 0.3694        & 0.4957           & \underline{0.6190}         & \textbf{0.6450} &  4.20\%  \\ 
									                  & \textbf{NDCG@20}                 & 0.1325          & 0.1504       & 0.1582            & 0.1588         & 0.1966        & 0.1790         & 0.1944        & 0.2989           & \underline{0.3225}         & \textbf{0.3533} &  9.55\%  \\ 
									                  & \textbf{HR@50}                   & 0.2652          & 0.2934       & 0.3658            & 0.4873         & 0.5227        & 0.5045         & 0.4607        & 0.6036           & \underline{0.7685}         & \textbf{0.7749} &  0.83\% \\ 
									                  & \textbf{NDCG@50}                 & 0.1417          & 0.1599       & 0.1747            & 0.1891         & 0.2198        & 0.1984         & 0.2029        & 0.3203           & \underline{0.3535}         & \textbf{0.3804} &  7.61\% \\
	\bottomrule
	\end{tabular}
  }
 \vspace{-0.2cm}
\end{table*}

\subsubsection{\textbf{Baselines}}
To demonstrate the performance of our model, we compare our MB-HGCN with several representative recommendation models, including three single-behavior models and six multi-behavior models. 

\textbf{Single-behavior model:}

\begin{itemize}[leftmargin=*]
  \item \textbf{MF-BPR}~\cite{BPR}. BPR is a widely used optimization strategy, which assumes that the predicted scores of positive samples are higher than that of negative ones. MF-BPR has been widely used as a baseline to evaluate the performance of newly proposed models. 
  
  \item \textbf{NCF}~\cite{NMTR}. It is a representative model combining neural network and CF, which combines shallow generalized matrix factorization model and deep multi-layer perceptron model to learn the interaction between users and items.

  \item \textbf{LightGCN}~\cite{LightGCN}. It removes the feature transformation and nonlinear activation components in the standard GCN model, and only keeps the core neighborhood aggregation component, which simplifies the model structure and achieves a significant performance improvement over its counterpart.
\end{itemize}

\textbf{Multi-behavior model:}

\begin{itemize}[leftmargin=*]
  \item \textbf{R-GCN}~\cite{SchlichtkrullKB18}. R-GCN differentiates the relations between nodes via edge types in the graph and designs different propagation layers for different types of edges to model the relation information. This model can adapt to the multi-behavior recommendation.

  \item \textbf{NMTR}~\cite{NMTR}. It is a deep learning model for multi-behavior recommendation, which designs a neural network for each behavior. It sequentially passes the interaction score among behaviors and also adopts multi-task learning for joint optimization.

  \item \textbf{MBGCN}~\cite{MBGCN}. This model constructs a heterogeneous graph to learn user preferences through user-item propagation and adopts a linear aggregation for feature fusion. In addition, item-item propagation is exploited to enhance item embedding learning.

  \item \textbf{GNMR}~\cite{GNMR}. This model designs a relation aggregation network to model interaction heterogeneity and attempts to explore the dependencies among different types of behaviors via recursive embedding propagation over the heterogeneous graph.

  \item \textbf{S-MBRec}~\cite{SMBREC}. This model consists of a supervised and a self-supervised learning task, which separately learns the user and item embeddings from each behavior and adopts a star-style contrastive learning strategy to construct a contrastive view pair for the target and each auxiliary behavior.
  
  \item \textbf{CRGCN}~\cite{CRGCN}. This model designs a cascaded residual network to explore the connection between different behaviors from the perspective of embedding propagation. The multi-task learning is also adopted for joint optimization.

\end{itemize}

\subsubsection{\textbf{Parameter Settings}}

Our model is implemented by Pytorch\footnote{https://pytorch.org/}.
In the implementation of all methods, the mini-batch size and embedding size are set to 1024 and 64, respectively~\cite{LightGCN}. Adam~\cite{Adam} optimizer is adopted for the optimization. In addition, we employ grid search to tune the learning rate and regularization weights (\textit{i.e.}, $\beta$) in the $[1e^{-2}, 3e^{-3}, 1e^{-3}, 1e^{-4}]$ and $[1e^{-2}, 1e^{-3}, 3e^{-4}, 1e^{-4}]$ ranges, respectively. Meanwhile, we carefully tune the hyperparameters in the baselines according to their original papers, and an early stop strategy is adopted in the training stage.

\subsection{Overall Performance (RQ1)}
In this section, we report the performance comparisons between our MB-HGCN model and all the baselines. The results on the three datasets are shown in Table~\ref{tab:overall}. Overall, the performance of multi-behavior methods outperforms that of single-behavior methods, which demonstrates the effectiveness of exploiting multiple behaviors. Among the multi-behavior methods, our MB-HGCN significantly outperforms other multi-behavior methods. Comparing with the best baseline, the average improvement  of \textit{HR@K} and \textit{NDCG@K} across \textit{top-K} for $(K= {10, 20, 50})$ are 66.41\% and 69.85\% on Tmall dataset, 11.12\% and 11.97\% on Beibei dataset, 3.92\% and 9.43\% on Jdata dataset, respectively. This is a remarkable improvement in the recommendation accuracy,  demonstrating the superiority of our model.

Among the single-behavior methods, NCF generally outperforms MF-BPR due to its ability to model the complex and nonlinear relationships between user-item interactions using a neural network architecture. However, LightGCN exhibits the best performance among single-behavior methods. This result confirms the effectiveness of GCN-based approaches in capturing the user-item interactions information, as LightGCN uses a simplified GCN model that emphasizes the importance of neighborhood aggregation. The superior performance of LightGCN highlights the importance of leveraging graph-based modeling techniques for recommendation tasks.
Among the multi-behavior methods, R-GCN, which directly combines embeddings learned separately from each behavior with a simple summation, exhibits poor performance, and in some cases, even performs worse than the single-behavior method LightGCN. This suggests that the straightforward aggregation of auxiliary behavior embeddings may have detrimental effects on recommendation accuracy. 
In contrast, MBGCN and GNMR adopt alternative strategies for embedding aggregation, and both achieve superior performance compared to R-GCN, which validates that different behaviors contribute differently to the target behavior.
Moreover, NMTR and CRGCN consider the relationships among multi-behaviors through cascading modeling and both yield better performance than the aforementioned methods.
NMTR models the cascading effects indirectly through interaction scores of different behaviors. In contrast, CRGCN directly incorporates cascading  effects into the embedding learning process, leading to superior performance over NMTR.
CRGCN is also the best-performing baseline in our experiments, leveraging multi-behavior relationships in embedding learning. However, MB-HGCN can outperform CRGCN by a large margin, mainly due to its hierarchical learning strategy and aggregation strategies. Our ablation studies provide further insights into the effectiveness of different components in MB-HGCN.

It is worth noting that the improvement achieved on Tmall far exceeded that of the other two datasets. The primary reason for this substantial gap can be attributed to the greater variety of behavioral interactions among the different datasets. In comparison to Jdata, Tmall's \textit{collect} behavior yielded a comparable amount of data as the \textit{buy} behavior, which provided rich information. By comparison, the Beibei platform requires users to follow a strict sequence of behavior for making purchases, \textit{i.e.},  \textit{view}$\to$\textit{cart}$\to$\textit{buy}. Consequently, the global embedding learned in our model reflected the \textit{view} behavior, which limits the performance of our model.

\subsection{Ablation Study (RQ2)}
In this section, we conduct extensive ablation studies to examine the validity of different components in our model.

\subsubsection{\textbf{Effect of the embedding learning in graph $\mathcal{G}$}} 
We design a  hierarchical graph convolutional network for embedding learning, where a unified graph $\mathcal{G}$ is utilized to learn a coarse-grained global embedding, which is then used as a shared initialization for refining embeddings in behavior-specific graphs. To validate the effectiveness of learning coarse-grained global embeddings, we conduct an experiment where we remove the unified graph component and compare the results to the original model that retain the unified graph. Specifically, we train the model without the unified graph $\mathcal{G}$ and leverage the initialized embedding (\textit{i.e.}, $\bm{e_u}^0$ and $\bm{e_i}^0$) as the initialization of the behavior-specific graph $\mathcal{G}_k$ ($k \in [1, K]$). Experimental results are reported in Table~\ref{tab:cel}.


\begin{table}[hbt]
  \caption{Effect of the embedding learning in graph $\mathcal{G}$ (\textit{w. $\mathcal{G}$} and \textit{w/o. $\mathcal{G}$} represent with and without embedding learning in graph $\mathcal{G}$, respectively).}
  \label{tab:cel}
  \resizebox{\columnwidth}{!}{
	\begin{tabular}{rcccccc}
	\toprule
	\multirow{2}{*}{\textbf{Method}} & \multicolumn{2}{c}{\textbf{Tmall}} & \multicolumn{2}{c}{\textbf{Beibei}} & \multicolumn{2}{c}{\textbf{Jdata}} \\ \cmidrule(lr){2-3} \cmidrule(lr){4-5} \cmidrule(l){6-7}
									 & \textbf{HR@10}  & \textbf{NDCG@10} & \textbf{HR@10}  & \textbf{NDCG@10}  & \textbf{HR@10}  & \textbf{NDCG@10} \\ \hline
	\textbf{\textit{w/o. $\mathcal{G}$}}                & 0.0394          & 0.0198           & 0.0420          & 0.0213            & 0.2709          & 0.1608           \\
	\textbf{\textit{w. $\mathcal{G}$}}                  & \textbf{0.1461} & \textbf{0.0770}  & \textbf{0.0619}  & \textbf{0.0297}  & \textbf{0.5338} & \textbf{0.3238}  \\ 
	\bottomrule
	\end{tabular}
  }
\end{table}

The experimental results suggest that removing the unified graph component leads to a significant decrease in performance. This result is attributed to the coarse-grained global embeddings learned in the unified graph component can provide better initialization for refining embeddings in behavior-specific graphs, which allows for more accurate learning in those graphs. Moreover, it is observed that there is a tremendous performance difference between the two models with and without the unified graph. In fact, without the unified graph, the model degenerates to a variant of R-GCN, where the difference is the embedding aggregation strategy. Compared with the results in Table~\ref{tab:overall}, the model \textit{w/o. $\mathcal{G}$} significantly outperformed R-GCN. The results provide evidence for the effectiveness of the proposed embedding aggregation strategies for users and items, and further verifies the validity of the unified graph design.


\subsubsection{\textbf{Effect of the user embedding aggregation strategy}} \label{user_agg}
Our intuition for designing user embedding aggregation strategies is that user interests vary across behaviors, and thus, not all user preferences contribute to the prediction of the target behavior. Therefore, we design a simple adaptive embedding aggregation strategy for user embeddings. To verify the effectiveness of the design for adaptive user embedding aggregation strategy, we conduct three experiments: 1) \textit{sum agg.}, we remove our adaptive user embedding aggregation module and directly sum different behavior-specific embeddings for information aggregation. 2) \textit{linear agg.}, we replace our adaptive user embedding aggregation module with linear aggregation, which assigns different weights based on the number of interactions for each behavior (the same to the item aggregation strategy). 3) \textit{adaptive agg.}, our proposed adaptive embedding aggregation strategy. Experimental results are reported in Table~\ref{tab:user_agg}.
\begin{table}[hbt]
  \caption{Comparison of three different user embedding aggregation strategies.}
  \label{tab:user_agg}
  \resizebox{\columnwidth}{!}{
	\begin{tabular}{rcccccc}
	\toprule
	\multirow{2}{*}{\textbf{Method}} & \multicolumn{2}{c}{\textbf{Tmall}} & \multicolumn{2}{c}{\textbf{Beibei}} & \multicolumn{2}{c}{\textbf{Jdata}} \\ \cmidrule(lr){2-3} \cmidrule(lr){4-5} \cmidrule(l){6-7}
									 & \textbf{HR@10}  & \textbf{NDCG@10} & \textbf{HR@10}  & \textbf{NDCG@10}  & \textbf{HR@10}  & \textbf{NDCG@10} \\ \hline
  \textbf{\textit{sum agg.}}                 & 0.0971                     & 0.0488                     & 0.0516                     & 0.0257                     & 0.4068                     & 0.2356                     \\
  \textbf{\textit{linear agg.}}             & 0.1263                     & 0.0687                     & 0.0575                     & 0.0288                     & 0.4888                     & 0.2935 \\

  \textbf{\textit{adaptive agg.}}                   & \textbf{0.1461}            & \textbf{0.0770}            & \textbf{0.0619}            & \textbf{0.0297}            & \textbf{0.5338}            & \textbf{0.3238}  \\
	\bottomrule
	\end{tabular}
  }
\end{table}

The experimental results demonstrate that adopting the adaptive aggregation strategy achieves the best performance. The \textit{sum agg.} method yields poor performance due to the varying interests that users exhibit in different behaviors, and the aggregation strategy lacks consideration of the importance of each behavior. Although the \textit{linear agg.} method considers the importance of different behaviors, behaviors with more interactions may not necessarily reflect more accurate user preferences. In contrast, our adaptive aggregation strategy aggregates relevant information at the feature level based on the similarity between different behaviors, resulting in better aggregation of relevant information. It is worth mentioning that our aggregation scheme does not introduce any additional parameters into the model. This avoids the potential risks of negative impact on the embedding learning process from additional parameters introduced by the aggregation scheme.

\begin{figure}[hbt]
  \centering
  \includegraphics[width=\columnwidth]{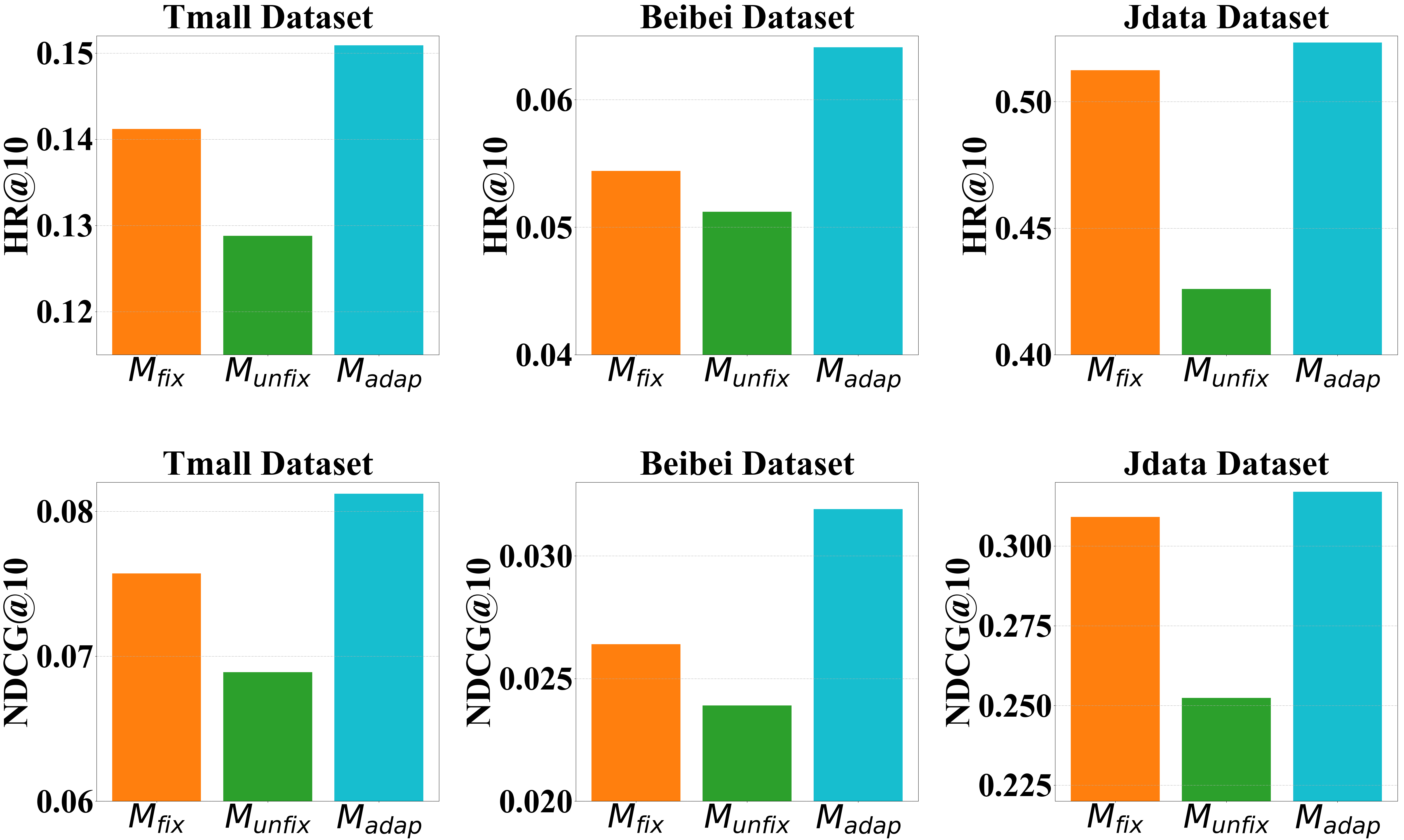}
  \caption{Impact of aggregation optimization.}
  \label{fig:impac}
  \vspace{-0.3cm}
\end{figure}

To verify this point, we perform an additional experiments. We pre-train our model to keep the optimal embeddings that learned from each behavior and remove the aggregation of global embeddings (\textit{i.e.}, the operation of Eq.~\ref{eq:fuse_u}) to eliminate the effects of global embeddings. The goal is to only retain the training of target behaviors to avoid the effects of multi-task learning. On this basis, we compare the performance of the following three variants: 1) $M_{unfix}$, which employs linear aggregation strategy for user embedding aggregation. 2) $M_{fix}$, which fixes the parameters in the embedding learning process based on the first experiment. 3) $M_{adap}$, which adopts our adaptive aggregation strategy for user embedding aggregation. The experimental results are reported in Fig.~\ref{fig:impac}.

According to Fig.~\ref{fig:impac}, we can observe that for the two linear aggregation methods, the one with fixed parameters significantly outperforms the one with unfixed parameters. This is because the supervision signal cannot be transferred to the embedding learning process in the method of fixed parameters, in which only the parameters of linear aggregation process are optimized. This suggests that optimization of aggregation parameters may lead to locally optimal solutions for embedding learning. Furthermore, the method that adopts the adaptive aggregation strategy is better than the two methods which employ linear aggregation. The reason is that our adaptive aggregation strategy does not introduce any parameters, facilitating the embedded learning to be optimized on the right direction.

\subsubsection{\textbf{Effect of the item embedding aggregation strategy}}
In this experiment, we evaluate the effectiveness of a linear aggregation strategy for item embeddings aggregation. Considering that behaviors with more interactions may reflect more comprehensive features of items, we assign weights to each behavior based on its number of interactions (as shown in Eq.~\ref{eq:weight}). To verify the effectiveness of this design, we conduct the following experiments: 1) \textit{fix $\gamma_{ik}$}, we assign the same weight (\textit{i.e.}, set $\gamma_{ik}=1$) to each behavior for embedding aggregation. 2) \textit{w/o. $w_k$}, we remove the learnable parameter $w_k$ and strictly assign weights based on the number of interactions for each behavior. 3) \textit{w. $w_k$}, which keeps the learnable parameter $w_k$ allows for fine-tuning the importance of different behaviors (our approach). The experimental results are reported in Table~\ref{tab:weight_sum}.


\begin{table}[hbt]
  \caption{Effect of the item aggregation strategy (\textit{w. $w_k$} and \textit{w/o. $w_k$} represent with and without $w_k$, respectively).}
  \label{tab:weight_sum}
  \resizebox{\columnwidth}{!}{
	\begin{tabular}{rcccccc}
	\toprule
	\multirow{2}{*}{\textbf{Method}} & \multicolumn{2}{c}{\textbf{Tmall}} & \multicolumn{2}{c}{\textbf{Beibei}} & \multicolumn{2}{c}{\textbf{Jdata}} \\ \cmidrule(lr){2-3} \cmidrule(lr){4-5} \cmidrule(l){6-7}
									 & \textbf{HR@10}  & \textbf{NDCG@10} & \textbf{HR@10}  & \textbf{NDCG@10}  & \textbf{HR@10}  & \textbf{NDCG@10} \\ \hline
  \textbf{\textit{fix $\gamma_{ik}$}}                & 0.1285          & 0.0686           & 0.0587           & 0.0276           & 0.4685          & 0.2795           \\
  \textbf{\textit{w/o. $w_k$}}                    & 0.1408          & 0.0762           & 0.0594           & 0.0304           & 0.4814          & 0.2906           \\
  
  \textbf{\textit{w. $w_k$}}        & \textbf{0.1461} & \textbf{0.0770}  & \textbf{0.0619}  & \textbf{0.0312}  & \textbf{0.5338} & \textbf{0.3238}  \\
	\bottomrule
	\end{tabular}
  }
\end{table}

The results in Table~\ref{tab:weight_sum} indicate a significant improvement for the weight allocation method (\textit{w/o. $w_k$}) over the non-weight allocation method (\textit{fix $\gamma_{ik}$}), which supports our viewpoint that behaviors with more interactions reflect more comprehensive item features. In the two weight allocation methods, \textit{w/o. $w_k$} and \textit{w. $w_k$}, the method that fine-tuned the weights using the learnable parameter achieved better performance, indicating that the contribution of different behaviors varies for different items. Therefore, fine-tuning the weights via the learnable parameter can better aggregate the representation of items, further validating the effectiveness of our proposed strategy.


\subsubsection{\textbf{Effect of the global embedding aggregation}} 
Aggregate global embeddings into the final embedding, as shown in Eq.~\ref{eq:fuse_u}, is to obtain more comprehensive representation. We conduct an ablation study to verify this point by comparing it with the variant without considering the global embeddings. The experimental results are reported in Table~\ref{tab:emb_sum}.

\begin{table}[hbt]
  \caption{Effect of the global embedding aggregation(\textit{w. c.g.} and \textit{w/o. c.g.} represent with and without global embedding aggregation, respectively).}
  \label{tab:emb_sum}
  \resizebox{\columnwidth}{!}{
	\begin{tabular}{rcccccc}
	\toprule
	\multirow{2}{*}{\textbf{Method}} & \multicolumn{2}{c}{\textbf{Tmall}} & \multicolumn{2}{c}{\textbf{Beibei}} & \multicolumn{2}{c}{\textbf{Jdata}} \\ \cmidrule(lr){2-3} \cmidrule(lr){4-5} \cmidrule(l){6-7}
									 & \textbf{HR@10}  & \textbf{NDCG@10} & \textbf{HR@10}  & \textbf{NDCG@10}  & \textbf{HR@10}  & \textbf{NDCG@10} \\ \hline
  \textbf{\textit{w/o. c.g.}}                & 0.1257          & 0.0665           & 0.0579           & 0.0271           & 0.4606          & 0.2739           \\
  \textbf{\textit{w. c.g.}}                  & \textbf{0.1461} & \textbf{0.0770}  & \textbf{0.0619}  & \textbf{0.0297}  & \textbf{0.5338} & \textbf{0.3238}   \\
	\bottomrule
	\end{tabular}
  }
\end{table}

It is shown that the method with global embedding aggregation, our model can gain a relative improvement of 16.23\% and 15.79\% on Tmall, 6.91\% and 9.59\% on Beibei, 15.89\% and 18.22\% on Jdata for \emph{HR@10} and \emph{NDCG@10}, respectively. This demonstrates that aggregate global embedding can indeed improve performance. Global embeddings reflect coarse-grained user preferences, while behavior-specific embeddings reflect fine-grained user preferences. Combining these two types of embeddings can provide a comprehensive and hierarchical representation of user preferences, which can further improve recommendation performance. It again justifies the effectiveness of our hierarchical design by learning embedding from both global and behavior-specific levels.

\subsubsection{\textbf{Effect of multi-task learning}} 
We adopt a multi-task learning (MTL) framework for joint optimization. To verify its effectiveness, we compare the method with and without MTL, in which the method without multi-task learning train the target behavior in a single-task. The experimental results are reported in Table~\ref{tab:mtl}.

\begin{table}[hbt]
  \caption{Effect of the joint optimization(\textit{w. MTL} and \textit{w/o. MTL} represent with and without MTL, respectively).}
  \label{tab:mtl}
  \resizebox{\columnwidth}{!}{
	\begin{tabular}{rcccccc}
	\toprule
	\multirow{2}{*}{\textbf{Method}} & \multicolumn{2}{c}{\textbf{Tmall}} & \multicolumn{2}{c}{\textbf{Beibei}} & \multicolumn{2}{c}{\textbf{Jdata}} \\ \cmidrule(lr){2-3} \cmidrule(lr){4-5} \cmidrule(l){6-7}
									 & \textbf{HR@10}  & \textbf{NDCG@10} & \textbf{HR@10}  & \textbf{NDCG@10}  & \textbf{HR@10}  & \textbf{NDCG@10} \\ \hline
  \textbf{\textit{w/o. MTL}}                 & 0.1393          & 0.0660           & 0.0385           & 0.0182           & 0.4428          & 0.2615           \\
  \textbf{\textit{w. MTL}}                  & \textbf{0.1461} & \textbf{0.0770}  & \textbf{0.0619}  & \textbf{0.0297}  & \textbf{0.5338} & \textbf{0.3238}   \\
	\bottomrule
	\end{tabular}
  }
\end{table}

The experimental results, reported in Table~\ref{tab:mtl}, demonstrate that the MTL method outperforms the single-task method across all three datasets, indicating the effectiveness of MTL. The underlying reason for its effectiveness lies in the fact that our model treats each behavior as an independent task during training, and the better the behavior-specific embedding fits the user preferences exhibited in current behavior during the training process, the more accurately relevant information can be aggregated in the adaptive aggregation stage, leading to more precise predictions. It is worth noting that the scale of interaction data for different behaviors may affect the performance of multi-task learning. Therefore, considering the importance of different behaviors is necessary, but it is not the focus of our current research. We plan to research it in future work.



\subsection{GCN layer Study (RQ3)}
\begin{figure}[hbt]
  \centering
  \includegraphics[width=\columnwidth]{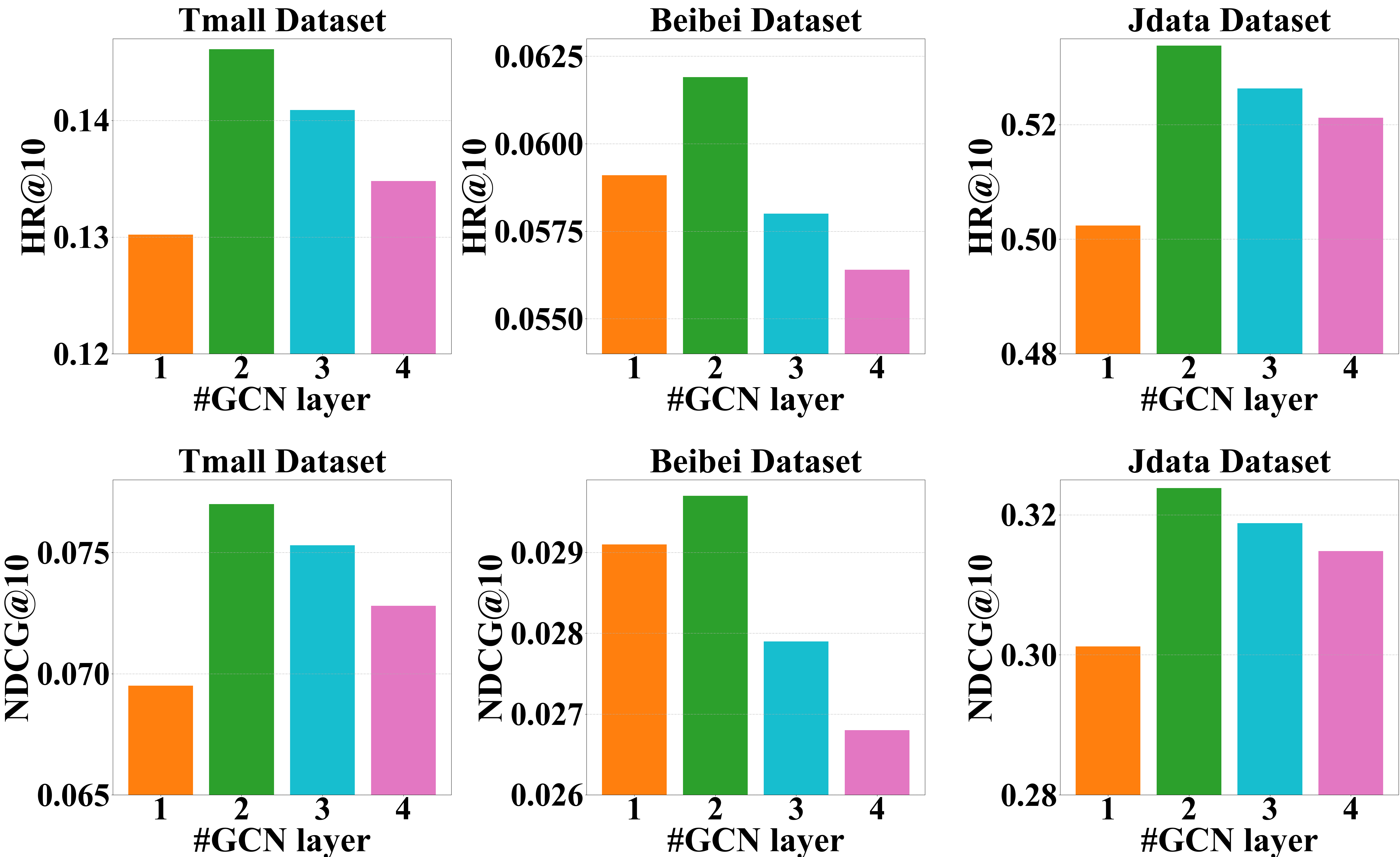}
  \caption{Performance comparison at different GCN layers.}
  \label{fig:gcn_layer}

\end{figure}

Our model adopts LightGCN as the backbone to perform convolution operations in each graph. From the perspective of the overall structure, the convolution operations on graph $\mathcal{G}$ and graph $\mathcal{G}_k$ successively are similar to simply increasing the number of GCN layers. To this end, we compare the effects of GCN layers with different number settings. The results are reported in Fig.~\ref{fig:gcn_layer}

From Fig.~\ref{fig:gcn_layer}, it can be seen that with the number of GCN layers increases, the performance will first increase with the increasing number of GCN layers and then drops when stacking more layers, which is consistent with the results observed in single-behavior methods LightGCN~\cite{LightGCN} and NGCF~\cite{NGCF}.  The best performance is obtained when the number of GCN layers is 2 in our experiments. 

\subsection{Cold-start Problem (RQ4)}
The cold-start problem in recommender systems refers to the situation where a new user or item is added to the system, and there is insufficient historical data available to provide personalized recommendations. Multi-behavior recommendation is one approach to alleviate the cold-start problem by considering multiple behaviors data. Such behavior data may include rich information that can help better understand user preferences. In this section, we will verify the capabilities of our model to tackle this problem.
\begin{figure}[hbt]
  \centering
  \includegraphics[width=\columnwidth]{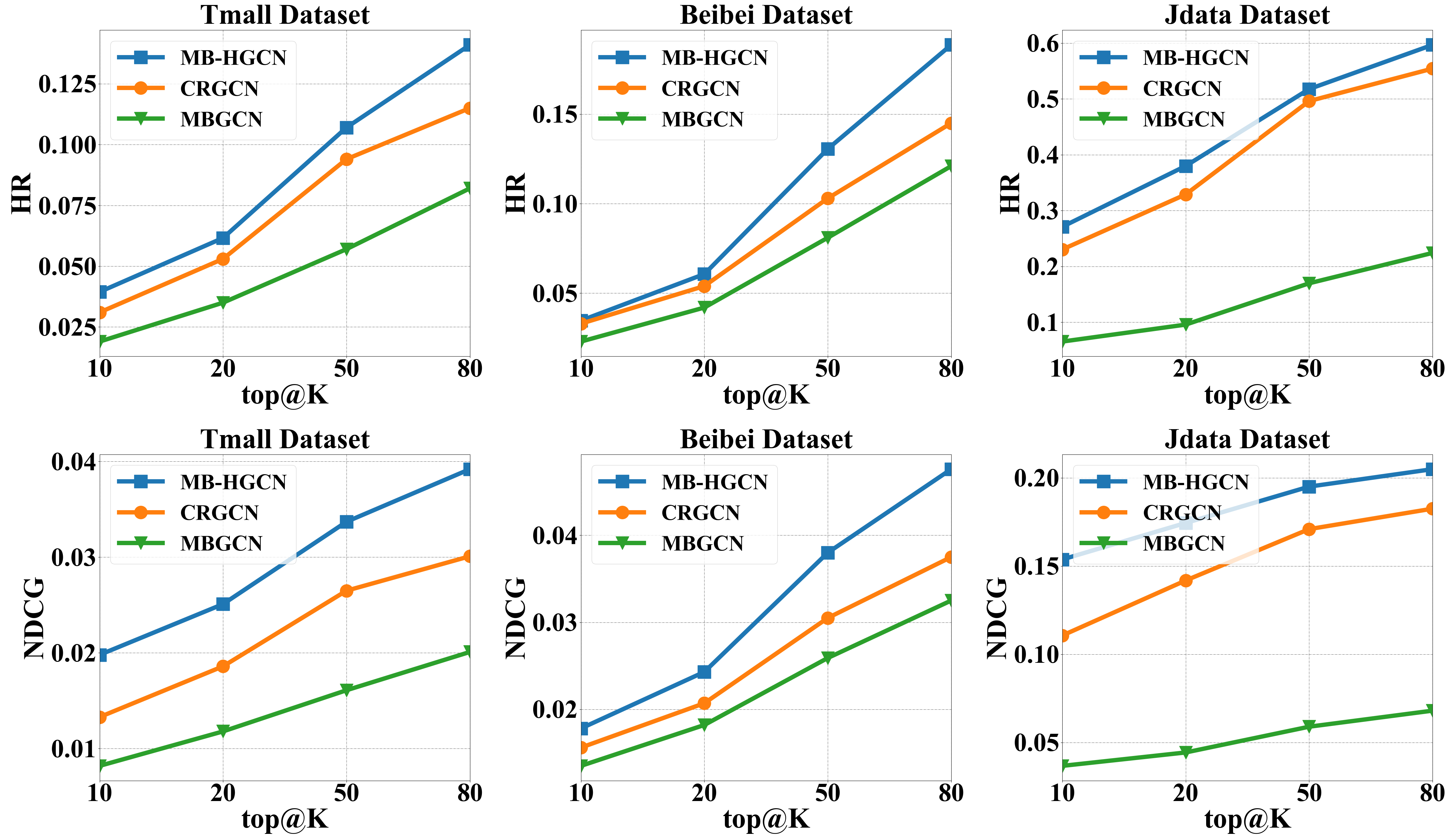}
  \caption{Performance comparisons among MB-HGCN, CRGCN, and MBGCN for cold-start users.}
  \label{fig:cold_start}

\end{figure}
We compare our MB-HGCN with two models, CRGCN and MBGCN, where CRGCN is the best baseline, and MBGCN designs a item-based scoring module to alleviate the cold start problem. To perform the study, we follow previous work~\cite{CRGCN, MBGCN} to randomly select 1,000 users from the test set as cold-start users and remove all of their \textit{buy} behavior records from the training set. In addition, for other behaviors in the training set, we also remove the user-item pairs involved in buy behavior. These 1,000 users are simulated as the hard cold-start users with no buy behavior records. This process ensures that these 1000 users do not have any prior preference information about items, thus simulating them as hardcore cold-start users with no \textit{buy} behavior records. We then train the model with the remaining records using the settings described in ~\ref{Experiment Settings}. Finally, use the trained model to provide personalized recommendations for these 1,000 cold-start users.

The experimental results are shown in Fig.~\ref{fig:cold_start}. It can be observed that out MB-HGCN consistently outperforms CRGCN and MBGCN across all three datasets. Compared with CRGCN, the average improvement of our model are 19.94\% and 35.30\% on Tmall dataset, 18.62\% and 20.75\% on Beibei dataset and 11.28\% and 22.08\% on Jdata dataset in terms of HR@K and NDCG@K. This experimental result indicates that our model is able to better utilize multi-behavioral data to learn user preferences for the target behavior recommendation. This should be attributed to the design of hierarchical graph convolutional network, we learn user preferences from a coarse-grained global level to a fine-grained behavior-specific level in this design. Therefore, even if users do not have \textit{buy} behavior, our model can still learn coarse-grained user preferences for the target behavior recommendation. In contrast, the sequential modeling of CRGCN fails to effectively learn random behaviors such as \textit{collect} behavior that are uncertain to occur, resulting in suboptimal results. In addition, CRGCN shows a significant improvement compared to MBGCN due to its cascading design, which can effectively utilize the effect of cascading behaviors to refine user preferences, while the weighted aggregation strategy adopted by MBGCN may not be able to capture the complex interrelationships between behaviors.

\subsection{Embedding Learning Analysis (RQ5)}
In recommender systems, embeddings are commonly used to represent users.  Each position in the embedding can be viewed as a potential interest feature for the user~\cite{KorenBV09, HuKV08, Cheng0HSCAACCIA16}, and these interest features collectively form the user's preferences. In our model, we first learn user preferences from a coarse-grained global level to a fine-grained behavior-specific level, and then adaptively aggregate relevant information from auxiliary behaviors based on their similarities. To explore the changes in user interests during this process, we visualize the user embeddings during the process. In this visualization, the darkness of color represents the importance of the feature, with darker colors indicating greater importance. The sum of all feature values in the embedding equals 1. Specifically, we randomly select one user from each of the Tmall, Beibei, and Jdata datasets, and display the first 8 positions of their global embeddings, behavior-specific embeddings and the final embeddings which used for the target behavior recommendation.

\begin{figure}[hbt]
  \centering
  \includegraphics[width=\columnwidth]{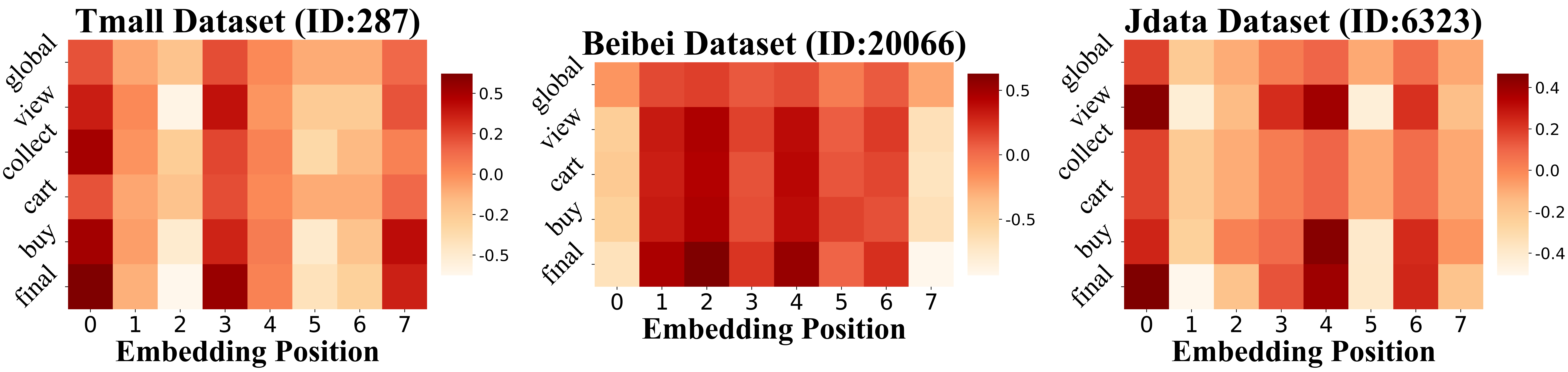}
  \caption{User embedding learning analysis. The y-axis displays the global embedding (\textit{global}), the behavior-specific embedding (\textit{i.e.}, \textit{view}, \textit{cart}, \textit{collect} and \textit{buy}) and the user embedding (\textit{final}) finally used for the target behavior recommendation, and the x-axis shows the feature information of the first 8 positions.}
  \label{fig:heap_map}

\end{figure}

The experimental results are shown in Fig.~\ref{fig:heap_map}. Overall, the interest distribution exhibited by the global embeddings is relatively evenly distributed across all three datasets. Compare with the global embedding, the feature values of behavior-specific embeddings exhibit different changes. Taking the \textit{view} behavior of Tmall dataset as an example, compared to the global embedding, the 0th, 1st, 3rd and 7th features in the view behavior-specific embedding have a darker color, indicating that users pay more attention to these features during browsing behavior. On the other hand, the 2nd, 4th, 5th and 6th features have a lighter color, suggesting that these features contribute less to user browsing behavior. This demonstrates that behavior-specific embedding refine and enhance the global embedding. Another interesting observation is that behavior-specific embeddings augment the degree of interest/disinterest (with darker colors becoming even darker and lighter colors becoming even lighter), without changing the properties of the feature (interested in becoming disinterested). This indicates that the global embedding can indeed represent users' coarse-grained preferences and further confirms that the behavior-specific embedding locally refines the global embedding. In addition, we observe that some behavioral features are consistent with global features (\textit{cart} in Tmall dataset and \textit{collect} and \textit{cart} in Jdata dataset), which is due to the lack of user-item interaction records in the corresponding behaviors. It also confirms that MB-HGCN can address the cold-start problem to some extent, \textit{i.e.}, when there is no \textit{buy} behavior, the model would retain the global embedding for recommendation. Finally, we adopt an adaptive aggregation strategy to obtain the final embedding for the target behavioral recommendation,  which is obtained by aggregating features based on \textit{buy} behavior. Taking the Tmall dataset as an example, the 0th and 3rd features in the final embedding are enhanced, while the 7th feature is slightly weakened, and other features with lower levels of interest (such as the 2nd, 5th, and 6th features) also adjusted to  some extent. This result also validates the effectiveness of the adaptive aggregation strategy we proposed.




%% file: 05_conclusion.tex
\section{conclusion} \label{conclusion}
In this work, we present a novel multi-behavior recommendation model named MB-HGCN, which can effectively exploit the multi-behavior information to learn user and item embeddings. In particular, a hierarchical graph network is designed to learn user preference from global to behavior-specific level. Moreover, two different aggregation strategies are applied to aggregate user and item embeddings learned from different behaviors. Extensive experimental results on three real-world benchmark datasets demonstrate the superiority of our model over the state-of-the-art MBR models. Further ablation studies verify the effectiveness of different components in our model. In the future, we plan to explore the relations among multi-behavior interactions in the embedding learning process, and conduct experiments on online systems with A/B testing to evaluate the performance of our proposed model.


%% file: main.bbl
\begin{thebibliography}{10}
\providecommand{\url}[1]{#1}
\csname url@samestyle\endcsname
\providecommand{\newblock}{\relax}
\providecommand{\bibinfo}[2]{#2}
\providecommand{\BIBentrySTDinterwordspacing}{\spaceskip=0pt\relax}
\providecommand{\BIBentryALTinterwordstretchfactor}{4}
\providecommand{\BIBentryALTinterwordspacing}{\spaceskip=\fontdimen2\font plus
\BIBentryALTinterwordstretchfactor\fontdimen3\font minus
  \fontdimen4\font\relax}
\providecommand{\BIBforeignlanguage}[2]{{%
\expandafter\ifx\csname l@#1\endcsname\relax
\typeout{** WARNING: IEEEtran.bst: No hyphenation pattern has been}%
\typeout{** loaded for the language `#1'. Using the pattern for}%
\typeout{** the default language instead.}%
\else
\language=\csname l@#1\endcsname
\fi
#2}}
\providecommand{\BIBdecl}{\relax}
\BIBdecl

\bibitem{zhang2019deep}
S.~Zhang, L.~Yao, A.~Sun, and Y.~Tay, ``Deep learning based recommender system:
  A survey and new perspectives,'' \emph{ACM Comput. Surv.}, vol.~52, no.~1,
  pp. 1--38, 2019.

\bibitem{Cheng2019mmalfm}
Z.~Cheng, X.~Chang, L.~Zhu, R.~C. Kanjirathinkal, and M.~S. Kankanhalli,
  ``{MMALFM:} explainable recommendation by leveraging reviews and images,''
  \emph{{ACM} Trans. Inf. Syst.}, vol.~37, no.~2, pp. 16:1--16:28, 2019.

\bibitem{KorenBV09}
Y.~Koren, R.~M. Bell, and C.~Volinsky, ``Matrix factorization techniques for
  recommender systems,'' \emph{Computer}, vol.~42, no.~8, pp. 30--37, 2009.

\bibitem{XuZCLS20}
Y.~Xu, L.~Zhu, Z.~Cheng, J.~Li, and J.~Sun, ``Multi-feature discrete
  collaborative filtering for fast cold-start recommendation,'' in
  \emph{Proceedings of the 34th {AAAI} Conference on Artificial
  Intelligence}.\hskip 1em plus 0.5em minus 0.4em\relax {AAAI} Press, 2020, pp.
  270--278.

\bibitem{Koren08}
Y.~Koren, ``Factorization meets the neighborhood: {A} multifaceted
  collaborative filtering model,'' in \emph{Proceedings of the 14th {ACM}
  {SIGKDD} International Conference on Knowledge Discovery and Data
  Mining}.\hskip 1em plus 0.5em minus 0.4em\relax {ACM}, 2008, pp. 426--434.

\bibitem{NingK11}
X.~Ning and G.~Karypis, ``{SLIM:} sparse linear methods for top-n recommender
  systems,'' in \emph{Proceedings of the 11th {IEEE} International Conference
  on Data Mining}.\hskip 1em plus 0.5em minus 0.4em\relax {IEEE}, 2011, pp.
  497--506.

\bibitem{BPR}
S.~Rendle, C.~Freudenthaler, Z.~Gantner, and L.~Schmidt{-}Thieme, ``{BPR:}
  bayesian personalized ranking from implicit feedback,'' in \emph{Proceedings
  of the 25th International Conference on Uncertainty in Artificial
  Intelligence}.\hskip 1em plus 0.5em minus 0.4em\relax {AUAI}, 2009, pp.
  452--461.

\bibitem{KabburNK13}
S.~Kabbur, X.~Ning, and G.~Karypis, ``{FISM:} factored item similarity models
  for top-n recommender systems,'' in \emph{Proceedings of the 19th {ACM}
  {SIGKDD} International Conference on Knowledge Discovery and Data
  Mining}.\hskip 1em plus 0.5em minus 0.4em\relax {ACM}, 2013, pp. 659--667.

\bibitem{PMF}
R.~Salakhutdinov and A.~Mnih, ``Probabilistic matrix factorization,'' in
  \emph{Proceedings of the 21st Annual Conference on Neural Information
  Processing Systems}.\hskip 1em plus 0.5em minus 0.4em\relax MIT Press, 2007,
  pp. 1257--1264.

\bibitem{NCF}
X.~He, L.~Liao, H.~Zhang, L.~Nie, X.~Hu, and T.~Chua, ``Neural collaborative
  filtering,'' in \emph{Proceedings of the 26th International Conference on
  World Wide Web}.\hskip 1em plus 0.5em minus 0.4em\relax {ACM}, 2017, pp.
  173--182.

\bibitem{WD}
H.~Cheng, L.~Koc, J.~Harmsen, T.~Shaked, T.~Chandra, H.~Aradhye, G.~Anderson,
  G.~Corrado, W.~Chai, M.~Ispir, R.~Anil, Z.~Haque, L.~Hong, V.~Jain, X.~Liu,
  and H.~Shah, ``Wide {\&} deep learning for recommender systems,'' in
  \emph{Proceedings of the 1st Workshop on Deep Learning for Recommender
  Systems}.\hskip 1em plus 0.5em minus 0.4em\relax {ACM}, 2016, pp. 7--10.

\bibitem{LightGCN}
X.~He, K.~Deng, X.~Wang, Y.~Li, Y.~Zhang, and M.~Wang, ``Lightgcn: Simplifying
  and powering graph convolution network for recommendation,'' in
  \emph{Proceedings of the 43rd International {ACM} {SIGIR} Conference on
  Research and Development in Information Retrieval}.\hskip 1em plus 0.5em
  minus 0.4em\relax {ACM}, 2020, pp. 639--648.

\bibitem{KGAT}
X.~Wang, X.~He, Y.~Cao, M.~Liu, and T.~Chua, ``{KGAT:} knowledge graph
  attention network for recommendation,'' in \emph{Proceedings of the 25th
  {ACM} {SIGKDD} International Conference on Knowledge Discovery and Data
  Mining}.\hskip 1em plus 0.5em minus 0.4em\relax {ACM}, 2019, pp. 950--958.

\bibitem{NMTR}
C.~Gao, X.~He, D.~Gan, X.~Chen, F.~Feng, Y.~Li, T.~Chua, L.~Yao, Y.~Song, and
  D.~Jin, ``Learning to recommend with multiple cascading behaviors,''
  \emph{{IEEE} Trans. Knowl. Data Eng.}, vol.~33, no.~6, pp. 2588--2601, 2021.

\bibitem{MBGCN}
B.~Jin, C.~Gao, X.~He, D.~Jin, and Y.~Li, ``Multi-behavior recommendation with
  graph convolutional networks,'' in \emph{Proceedings of the 43rd
  International {ACM} {SIGIR} Conference on Research and Development in
  Information Retrieval}.\hskip 1em plus 0.5em minus 0.4em\relax {ACM}, 2020,
  pp. 659--668.

\bibitem{BPRH}
H.~Qiu, Y.~Liu, G.~Guo, Z.~Sun, J.~Zhang, and H.~T. Nguyen, ``{BPRH:} bayesian
  personalized ranking for heterogeneous implicit feedback,'' \emph{Inf. Sci.},
  vol. 453, pp. 80--98, 2018.

\bibitem{SchlichtkrullKB18}
M.~S. Schlichtkrull, T.~N. Kipf, P.~Bloem, R.~van~den Berg, I.~Titov, and
  M.~Welling, ``Modeling relational data with graph convolutional networks,''
  in \emph{Proceedings of the 15th European Semantic Web Conference}.\hskip 1em
  plus 0.5em minus 0.4em\relax Springer, 2018, pp. 593--607.

\bibitem{GNMR}
L.~Xia, C.~Huang, Y.~Xu, P.~Dai, M.~Lu, and L.~Bo, ``Multi-behavior enhanced
  recommendation with cross-interaction collaborative relation modeling,'' in
  \emph{Proceedings of the 37th {IEEE} International Conference on Data
  Engineering}.\hskip 1em plus 0.5em minus 0.4em\relax {IEEE}, 2021, pp.
  1931--1936.

\bibitem{ZhangMCX20}
W.~Zhang, J.~Mao, Y.~Cao, and C.~Xu, ``Multiplex graph neural networks for
  multi-behavior recommendation,'' in \emph{Proceedings of the 29th {ACM}
  International Conference on Information and Knowledge Management}.\hskip 1em
  plus 0.5em minus 0.4em\relax {ACM}, 2020, pp. 2313--2316.

\bibitem{CMF}
A.~P. Singh and G.~J. Gordon, ``Relational learning via collective matrix
  factorization,'' in \emph{Proceedings of the 14th {ACM} {SIGKDD}
  International Conference on Knowledge Discovery and Data Mining}.\hskip 1em
  plus 0.5em minus 0.4em\relax {ACM}, 2008, pp. 650--658.

\bibitem{ZhaoCHC15}
Z.~Zhao, Z.~Cheng, L.~Hong, and E.~H. Chi, ``Improving user topic interest
  profiles by behavior factorization,'' in \emph{Proceedings of the 24th
  International Conference on World Wide Web}.\hskip 1em plus 0.5em minus
  0.4em\relax {ACM}, 2015, pp. 1406--1416.

\bibitem{LoniPLH16}
B.~Loni, R.~Pagano, M.~A. Larson, and A.~Hanjalic, ``Bayesian personalized
  ranking with multi-channel user feedback,'' in \emph{Proceedings of the 10th
  {ACM} Conference on Recommender Systems}.\hskip 1em plus 0.5em minus
  0.4em\relax {ACM}, 2016, pp. 361--364.

\bibitem{DingY0QLCJY18}
J.~Ding, G.~Yu, X.~He, Y.~Quan, Y.~Li, T.~Chua, D.~Jin, and J.~Yu, ``Improving
  implicit recommender systems with view data,'' in \emph{Proceedings of the
  27th International Joint Conference on Artificial Intelligence}.\hskip 1em
  plus 0.5em minus 0.4em\relax ijcai.org, 2018, pp. 3343--3349.

\bibitem{GuoQTLMW17}
G.~Guo, H.~Qiu, Z.~Tan, Y.~Liu, J.~Ma, and X.~Wang, ``Resolving data sparsity
  by multi-type auxiliary implicit feedback for recommender systems,''
  \emph{Knowl. Based Syst.}, vol. 138, pp. 202--207, 2017.

\bibitem{MATN}
L.~Xia, C.~Huang, Y.~Xu, P.~Dai, B.~Zhang, and L.~Bo, ``Multiplex behavioral
  relation learning for recommendation via memory augmented transformer
  network,'' in \emph{Proceedings of the 43rd International {ACM} {SIGIR}
  Conference on Research and Development in Information Retrieval}.\hskip 1em
  plus 0.5em minus 0.4em\relax {ACM}, 2020, pp. 2397--2406.

\bibitem{MB-GMN}
L.~Xia, Y.~Xu, C.~Huang, P.~Dai, and L.~Bo, ``Graph meta network for
  multi-behavior recommendation,'' in \emph{Proceedings of the 44th
  International {ACM} {SIGIR} Conference on Research and Development in
  Information Retrieval}.\hskip 1em plus 0.5em minus 0.4em\relax {ACM}, 2021,
  pp. 757--766.

\bibitem{ZhuLLSLCWN23}
Y.~Zhu, Q.~Lin, H.~Lu, K.~Shi, D.~Liu, J.~Chambua, S.~Wan, and Z.~Niu,
  ``Recommending learning objects through attentive heterogeneous graph
  convolution and operation-aware neural network,'' \emph{{IEEE} Trans. Knowl.
  Data Eng.}, vol.~35, no.~4, pp. 4178--4189, 2023.

\bibitem{SMBREC}
S.~Gu, X.~Wang, C.~Shi, and D.~Xiao, ``Self-supervised graph neural networks
  for multi-behavior recommendation,'' in \emph{Proceedings of the 31st
  International Joint Conference on Artificial Intelligence}.\hskip 1em plus
  0.5em minus 0.4em\relax ijcai.org, 2022, pp. 2052--2058.

\bibitem{DIPN}
L.~Guo, L.~Hua, R.~Jia, B.~Zhao, X.~Wang, and B.~Cui, ``Buying or browsing?:
  Predicting real-time purchasing intent using attention-based deep network
  with multiple behavior,'' in \emph{Proceedings of the 25th {ACM} {SIGKDD}
  International Conference on Knowledge Discovery and Data Mining}.\hskip 1em
  plus 0.5em minus 0.4em\relax {ACM}, 2019, pp. 1984--1992.

\bibitem{CKML}
C.~Meng, Z.~Zhao, W.~Guo, Y.~Zhang, H.~Wu, C.~Gao, D.~Li, X.~Li, and R.~Tang,
  ``Coarse-to-fine knowledge-enhanced multi-interest learning framework for
  multi-behavior recommendation,'' \emph{CoRR}, vol. abs/2208.01849, pp. 1--11,
  2022.

\bibitem{GHCF}
C.~Chen, W.~Ma, M.~Zhang, Z.~Wang, X.~He, C.~Wang, Y.~Liu, and S.~Ma, ``Graph
  heterogeneous multi-relational recommendation,'' in \emph{Proceedings of the
  35th {AAAI} Conference on Artificial Intelligence}.\hskip 1em plus 0.5em
  minus 0.4em\relax {AAAI} Press, 2021, pp. 3958--3966.

\bibitem{CRGCN}
M.~Yan, Z.~Cheng, C.~Gao, J.~Sun, F.~Liu, F.~Sun, and H.~Li, ``Cascading
  residual graph convolutional network for multi-behavior recommendation,''
  \emph{ACM Trans. Inf. Syst.}, pp. 1--24, 2023.

\bibitem{WanM18}
M.~Wan and J.~J. McAuley, ``Item recommendation on monotonic behavior chains,''
  in \emph{Proceedings of the 12th {ACM} Conference on Recommender
  Systems}.\hskip 1em plus 0.5em minus 0.4em\relax {ACM}, 2018, pp. 86--94.

\bibitem{UltraGCN}
K.~Mao, J.~Zhu, X.~Xiao, B.~Lu, Z.~Wang, and X.~He, ``Ultragcn: Ultra
  simplification of graph convolutional networks for recommendation,'' in
  \emph{Proceedings of the 30th {ACM} International Conference on Information
  and Knowledge Management}.\hskip 1em plus 0.5em minus 0.4em\relax {ACM},
  2021, pp. 1253--1262.

\bibitem{SVD-GCN}
S.~Peng, K.~Sugiyama, and T.~Mine, ``{SVD-GCN:} {A} simplified graph
  convolution paradigm for recommendation,'' in \emph{Proceedings of the 31st
  {ACM} International Conference on Information and Knowledge
  Management}.\hskip 1em plus 0.5em minus 0.4em\relax {ACM}, 2022, pp.
  1625--1634.

\bibitem{PLE}
H.~Tang, J.~Liu, M.~Zhao, and X.~Gong, ``Progressive layered extraction
  {(PLE):} {A} novel multi-task learning {(MTL)} model for personalized
  recommendations,'' in \emph{Proceedings of the 14th {ACM} Conference on
  Recommender Systems}.\hskip 1em plus 0.5em minus 0.4em\relax {ACM}, 2020, pp.
  269--278.

\bibitem{NGCF}
X.~Wang, X.~He, M.~Wang, F.~Feng, and T.~Chua, ``Neural graph collaborative
  filtering,'' in \emph{Proceedings of the 42nd International {ACM} {SIGIR}
  Conference on Research and Development in Information Retrieval}.\hskip 1em
  plus 0.5em minus 0.4em\relax {ACM}, 2019, pp. 165--174.

\bibitem{GCMC}
R.~van~den Berg, T.~N. Kipf, and M.~Welling, ``Graph convolutional matrix
  completion,'' in \emph{Proceedings of the 24th {ACM} {SIGKDD} International
  Conference on Knowledge Discovery and Data Mining}.\hskip 1em plus 0.5em
  minus 0.4em\relax {ACM}, 2018, pp. 1--7.

\bibitem{Gao0GCFLCJ19}
C.~Gao, X.~He, D.~Gan, X.~Chen, F.~Feng, Y.~Li, T.~Chua, and D.~Jin, ``Neural
  multi-task recommendation from multi-behavior data,'' in \emph{Proceedings of
  the 35th {IEEE} International Conference on Data Engineering}.\hskip 1em plus
  0.5em minus 0.4em\relax {IEEE}, 2019, pp. 1554--1557.

\bibitem{HR}
G.~Karypis, ``Evaluation of item-based top-n recommendation algorithms,'' in
  \emph{Proceedings of the 10th {ACM} {CIKM} International Conference on
  Information and Knowledge Management}.\hskip 1em plus 0.5em minus 0.4em\relax
  {ACM}, 2001, pp. 247--254.

\bibitem{NDCG}
K.~J{\"{a}}rvelin and J.~Kek{\"{a}}l{\"{a}}inen, ``{IR} evaluation methods for
  retrieving highly relevant documents,'' in \emph{Proceedings of the 23rd
  Annual International {ACM} {SIGIR} Conference on Research and Development in
  Information Retrieval}, E.~J. Yannakoudakis, N.~J. Belkin, P.~Ingwersen, and
  M.~Leong, Eds.\hskip 1em plus 0.5em minus 0.4em\relax {ACM}, 2000, pp.
  41--48.

\bibitem{Adam}
D.~P. Kingma and J.~Ba, ``Adam: {A} method for stochastic optimization,'' in
  \emph{Proceedings of the 3rd International Conference on Learning
  Representations}.\hskip 1em plus 0.5em minus 0.4em\relax {OpenReview.net},
  2015, pp. 1--15.

\bibitem{HuKV08}
Y.~Hu, Y.~Koren, and C.~Volinsky, ``Collaborative filtering for implicit
  feedback datasets,'' in \emph{Proceedings of the 8th {IEEE} International
  Conference on Data Mining}.\hskip 1em plus 0.5em minus 0.4em\relax {IEEE}
  Computer Society, 2008, pp. 263--272.

\bibitem{Cheng0HSCAACCIA16}
H.~Cheng, L.~Koc, J.~Harmsen, T.~Shaked, T.~Chandra, H.~Aradhye, G.~Anderson,
  G.~Corrado, W.~Chai, M.~Ispir, R.~Anil, Z.~Haque, L.~Hong, V.~Jain, X.~Liu,
  and H.~Shah, ``Wide {\&} deep learning for recommender systems,'' in
  \emph{Proceedings of the 1st Workshop on Deep Learning for Recommender
  Systems}.\hskip 1em plus 0.5em minus 0.4em\relax {ACM}, 2016, pp. 7--10.

\end{thebibliography}
